\input epsf
\documentstyle[pre,aps]{revtex}
\draft
\begin{document}

\twocolumn[
\hsize\textwidth\columnwidth\hsize\csname @twocolumnfalse\endcsname

\date{\today}
\title{ THE ENERGY-SCALING APPROACH TO \\
 PHASE-ORDERING GROWTH LAWS}
\author{A. D. Rutenberg\cite{emailar} and A. J. Bray\cite{emailab}}
\address{Theoretical Physics Group \\
Department of Physics and Astronomy \\
The University of Manchester, M13 9PL, UK \\
cond-mat/9409088}
\maketitle
\widetext
\begin{abstract}
We present a simple, unified approach to determining the growth law for the
characteristic length scale, $L(t)$, in the phase ordering kinetics of a
system quenched from a disordered phase to within an ordered phase.
This approach, based on a scaling assumption for pair correlations,
determines $L(t)$ self-consistently for purely dissipative dynamics
by computing the time-dependence of the energy in two ways.
We derive growth laws for conserved and non-conserved $O(n)$ models,
including two-dimensional XY models and systems with textures. We
demonstrate that the growth laws for other systems, such as
liquid-crystals and Potts models, are determined by the type of topological
defect in the order parameter field that dominates the energy.
We also obtain generalized Porod laws for systems with topological textures.
\end{abstract}
\narrowtext
]

\section{Introduction}
\label{sec:intro}

The quench of a system from a disordered phase to an ordered phase
is a non-equilibrium process in which
energy is dissipated and topological defects, if present, are eliminated.
Typically, the system develops a scaling structure with a single length scale
that evolves in time as the various broken-symmetry
phases compete to select the ordered phase \cite{Gunton83,Langer92,Bray93e}.
In the thermodynamic limit, this length scale
will grow without bound, and the scaling structure will hold at late
times. When scaling holds, any theoretical or experimental
analysis of a system is simplified.  It is then natural to explore
the growth law of the single length scale, $L(t)$.
That is the aim of this paper, aspects of which have appeared
elsewhere \cite{Bray94,Rutenberg94a}.

Previous investigations of growth laws have been carried on a case-by-case
basis, and many of the predictions have been controversial. The approach
presented here is simple yet powerful, dealing with all
phase ordering systems within a common framework. This approach is
based on the role played
by topological defects in fixing the large-momentum behavior of two-point
correlation functions. The only restriction is that the dynamics be purely
dissipative, the major assumption that the scaling hypothesis is valid.
Even this assumption
may be relaxed: in systems with more than one characteristic
scale, our approach sets one relationship between the length scales.

Systems with scalar order parameters, such as
binary alloys and Ising models, have been well studied
\cite{Gunton83,Langer92,Bray93e,%
Onuki85a,Nagai83,Allen79,Rogers89,Kawakatsu85,Furukawa85,%
Hayakawa93a,Lifshitz62,Mason93,Shannon92,Roland88,%
Lifshitz61,Huse86,Gaulin87,Lee93}.
In such systems domains of both phases grow, and the intervening domain walls
(the characteristic topological defects of scalar systems)
decrease in total area and hence dissipate energy.
Recently there has been a growing interest in systems with vector and more
complex order parameters \cite{Bray93e}.
A series of experiments \cite{Wong92,Chuang91,Pargellis92}
and simulations
\cite{Yurke93,Newman90a,Mondello93,Rao94,%
Rutenberg94e,Mondello90,Blundell92,Blundell94,%
Zapotocky94,Toyoki93,Bray90,Toyoki91a,Mondello92,Toyoki91b,Siegert93,%
Siegert94,Puri94} have explored such systems.
A characteristic aspect of many of these systems
is the existence of stable topological defects with singular cores.  These
defects include domain walls, vortex lines, and points \cite{Toulouse76}.
If a system has a scaling structure, then the single length-scale $L(t)$ will
also characterize the curvature of any domain walls and vortex lines, and the
separation of any point defects.  One can
use the nature of the defects to determine the short distance
correlations in the system.  This connection between the topological
defects of the system and its correlations
can be exploited. The scaling assumption and the knowledge
of the defect structure are together sufficient to determine the growth law.

We calculate the growth law of the characteristic
length scale, {\em if scaling exists}, for
quenched systems with either scalar or vector fields,
with either short- or long-range attractive
interactions, and with or without conservation
laws. We do this by considering the time-dependence of the energy density
as the system relaxes towards a ground state.
First we calculate the energy density of the system.
Then we equate its time-derivative to the rate of energy-density dissipation
independently calculated from the local evolution of the order
parameter. From this we self-consistently determine the growth law, $L(t)$.
Our approach is independent
of the details of the system and of the initial conditions,
and hence reflects the observed universality of growth laws among
physical systems and simulations.

The symmetries of the system are reflected in the topological defects
seen in the system. The defect structure determines the
asymptotic behavior of correlations, so that
we do not need to use any approximation schemes in our approach.
For short-range interactions we find three regimes (shown in figure
\ref{FIG:SHORTRANGE}): defect dominated  scalar and XY [$O(2)$] systems both
with and without relevant conservation laws, and spin-wave dominated
\cite{magnetic} systems in which conservation laws are always relevant.
In the defect dominated regime topological defects
provide short-scale structure that determines
the energy density and the growth laws.
In the spin-wave dominated regime, topological defects, if present, do
not dominate the energetics or determine the growth law.
For all regimes, we find power-law growth laws which are independent of the
spatial dimension of the system, apart from special cases of one-dimensional
(1D) scalar systems and two-dimensional (2D) XY models. For the marginal cases
separating regimes we find novel logarithmic factors in the growth laws.

It is informative to contrast our approach with renormalization-group
(RG) work by Bray \cite{Bray89,Bray93b} (see also \cite{Bray93e,MC}).
In common is the universality with respect to the details of the
phase-ordering system, and the assumption of dynamical scaling.
There are three significant differences. The first is that the RG
approach only determines the dynamical exponent $z$, i.e. the
power law $1/z$ of the growth, but does not determine
any logarithmic factors. The second one is that the RG approach implicitly
assumes the existence of a simple fixed point
in which {\em all} correlation functions scale with a single length,
while our approach only requires the scaling of the two-time two-point
correlation function. The third difference is that
the RG approach is limited to systems with a relevant
conservation law, while our approach also applies with non-conserved dynamics.
In contrast to the RG methods, our ``Energy-Scaling'' approach
places the growth laws of both conserved and non-conserved scaling
systems on equivalent theoretical footing, and explicitly predicts
any logarithmic factors.  In principle, the results of this work could
be reached by the RG approach using
an explicit RG structure for non-equilibrium phase-ordering problems,
but, to our knowledge, that has not yet been developed.

In section \ref{sec:model} we introduce the model.
In section  \ref{sec:growth} we obtain growth laws for a variety of $O(n)$
systems. We introduce the scaling assumptions and then
calculate the energy density and the energy-density dissipation.
We also discuss systems with long-range attractive interactions.
These results have mainly been presented before \cite{Bray94},
though in a brief form (see also \cite{Rutenberg94a}).
In section \ref{sec:special} we discuss systems without topological
defects, then consider the collapse of isolated defect structures such as
spherical domains, 2D XY point defect pairs, and non-singular textures.
We use these results to obtain growth laws for 2D XY models, and also
asymptotic correlations and growth laws for systems with topological
textures. These have mostly {\em not} been presented
before and provide a detailed perspective on scalar, XY, and textured
phase-ordering systems. In section \ref{sec:review} we review relevant
work on phase-ordering growth laws.
We then discuss our results in section \ref{sec:discussion} and
summarize them in section \ref{sec:summary}.

\section{The Model}
\label{sec:model}

A generic energy functional for a phase-ordering system with
an $n$-component order parameter, $\vec{\phi}({\bf x})$,
and short-range interactions, is
\begin{equation}
\label{EQN:HAMILTONIAN}
H[\vec{\phi}] = \int d^d x\,\left[ (\nabla \vec{\phi})^2 + V(\vec{\phi})
\right]\ ,
\end{equation}
where $d$ is the spatial dimension and $V(\vec{\phi})$ is an isotropic,
``Mexican-hat'' shaped potential \cite{potential} such as
\begin{equation}
\label{EQN:POTENTIAL}
V(\vec{\phi})= V_0 ( \vec{\phi}^2 -1)^2.
\end{equation}
After a temperature quench into the ordered phase, the
equation of motion for the ordering kinetics of
the Fourier components $\vec{\phi}_{\bf k}$ \cite{Hohenberg77} is
\begin{equation}
\label{EQN:DYNAMICS}
\partial_t \vec{\phi}_{\bf k} =
-k^{\mu}\,(\delta H/\delta \vec{\phi}_{-{\bf k}}),
\end{equation}
where we only consider systems with purely dissipative dynamics.
We work at temperature $T=0$, with no thermal noise,
since $T$ is an ``irrelevant variable''
within the ordered phase \cite{Bray89,Newman90b}.
The conventional non-conserved model-A and conserved model-B dynamics
are $\mu=0$  and $\mu=2$, respectively. However, any $\mu>0$ enforces a
global conservation law for $\vec{\phi}$ \cite{Onuki85a}.

Shortly after the quench, the magnitude of the
order parameter nearly saturates and evolution
takes place by the motion of topological defects and, for vector systems,
by the relaxation of the director-field of the order parameter.
The role of the potential, $V(\vec{\phi})$,
in these later stages of evolution is different between
scalar and vector systems, though in neither case does it dominate
the gradient term in (\ref{EQN:HAMILTONIAN}).
Near defect cores, where the order parameter vanishes in order to reduce
the gradient energy, both the gradient and potential terms
contribute to the energy-density in proportion to the defect-core
density. Away from defect cores, the field is close to saturation
and varies over scales of order $L(t)$, the characteristic length.
Consider a small  region, away from defect cores, with a uniform gradient in
the
director field $\hat{\phi}$ and a uniform magnitude $|\vec{\phi}|= 1-\delta$.
The local energy density,
${\cal H} \equiv (\nabla \vec{\phi})^2 + V(\vec{\phi})$,
 will be of the same form for any potential $V(\vec{\phi})$
with a quadratic minimum:
\begin{equation}
\label{EQN:AMPEN}
{\cal H}[\vec{\phi}] \simeq (1-\delta)^2(\nabla \hat{\phi})^2 +4 V_0 \delta^2.
\end{equation}
The local energy density is then minimized if
\begin{equation}
\label{EQN:AMP}
\delta \simeq (\nabla \hat{\phi})^2 / (4 V_0) .
\end{equation}
Since $\nabla \hat{\phi} \sim 1/L(t)$, the contribution
to $\epsilon$ from the potential term, of order $\delta^2 \sim L^{-4}$,
is clearly subdominant for large $L(t)$. Including
non-uniform gradients in the director field, and
corresponding gradients of $\delta$, only contributes more subdominant terms.
For vector systems, the energy due to the
magnitude variations of the order parameter balance between the potential
and gradient terms, but both are dominated by the gradient term of the director
field.  For scalar systems, the director field is uniform away from defects
and so the energy density is dominated by the domain wall energy,
which is balanced between the potential and gradient terms. In both cases,
the gradient term is proportional to the local energy density. Thus the
scaling behavior of the average energy density of the system,
$\epsilon \equiv \langle {\cal H} \rangle$, is captured by
\begin{equation}
\label{EQN:EN}
\epsilon \sim \langle ( \nabla \vec{\phi})^2 \rangle \sim \int_{\bf k}
	k^2 \left< \vec{\phi}_{\bf k} \cdot \vec{\phi}_{-{\bf k}} \right>,
\end{equation}
where $\int_{\bf k}$ is the momentum integral $\int d^d k /(2\pi)^d$,
and the angle-brackets represent an average over initial conditions, or,
equivalently, a spatial average in the thermodynamic limit.
We independently calculate the rate of energy density dissipation,
$\dot{\epsilon} \equiv \partial_t \epsilon$,
by integrating the contribution from each Fourier mode:
\begin{eqnarray}
\label{EQN:ENDISS}
\dot{\epsilon} &=& \int_{\bf k} \left< (\delta H/\delta\vec{\phi}_{\bf k})
	\cdot \partial_t \vec{\phi}_{\bf k} \right>,  \nonumber \\
&=& - \int_{\bf k} k^{-\mu}\,
      	\left< \partial_t \vec{\phi}_{\bf k} \cdot
	\partial_t \vec{\phi}_{\bf -k}
	\right>\ ,
\end{eqnarray}
where we use the equation of motion (\ref{EQN:DYNAMICS}) to obtain
the second line.  We see the special role of the
two-point equal-time correlation function,
\begin{equation}
\label{EQN:CORR}
	S({\bf k},t) =
	\left< \vec{\phi}_{\bf k} \cdot \vec{\phi}_{-{\bf k}} \right>,
\end{equation}
which determines the energy density, and of the two-point
time-derivative correlation function,
\begin{equation}
\label{EQN:TTCORR}
        T({\bf k},t) \equiv
      	\left< \partial_t \vec{\phi}_{\bf k} \cdot
	\partial_t \vec{\phi}_{\bf -k}
	\right>\ ,
\end{equation}
which
determines the rate of energy-density dissipation. These correlation functions
will be the basis of our discussion of growth laws for systems that satisfy
dynamical scaling in the following section and subsequently
of isolated defects and growth laws in scalar, XY, and textured systems.
We will equate the time-derivative of $\epsilon$ from (\ref{EQN:EN}) to
$\dot{\epsilon}$ from (\ref{EQN:ENDISS}), and self-consistently determine
the growth-laws.

Before proceeding, we digress to discuss what can be learned from
dimensional analysis of the dynamics in Eq. (\ref{EQN:DYNAMICS}).
{}From the linear part, associated with the gradient term
in $H$, of the equation of motion (\ref{EQN:DYNAMICS})
we have the dimensional relation $[L] = [t^{1/(2+\mu)}]$. What other
length scales are there in the problem? The nonlinear part of
(\ref{EQN:DYNAMICS}), associated with the potential term in $H$,
gives $[L] = [\xi]$ where $\xi$, given by $V_0^{-1/2}$ for the potential
(\ref{EQN:POTENTIAL}), is the core size of any topological defects
present, e.g.\ the width of a domain wall. Finally, there is
the length scale $\xi_0$ associated with any short-range correlations
in the ensemble of initial conditions.
Thus dimensional analysis gives $[L] = [t^{1/(2+\mu)}] = [\xi] = [\xi_0]$,
implying
\begin{equation}
L(t) = t^{1/(2+\mu)} F(\xi^{2+\mu}/t,\,\xi_0^{2+\mu}/t)\ .
\label{EQN:DIMANAL}
\end{equation}
The growth law can only be changed from the naive (without $\xi$ or $\xi_0$)
dimensional result $L \sim t^{1/(2+\mu)}$ by non-trivial behavior of
$F$ at late times, when its arguments approach zero.
However, $\xi_0$ cannot enter into the growth law unless scaling is
violated (see, e.g. \cite{Newman90a,Rutenberg94e,Rutenberg94b}).
If $\xi_0$ did enter into the asymptotic growth law, then changing
the initial condition from the state at $t=0$ to one at $t_1 >0$
(with a larger $\xi_0$) will lead to a multiplicative change in the amplitude
of the growth law, rather than merely a shift of origin of the time coordinate.
This is unphysical, so any $\xi_0$ dependence implies a lack of scaling.
Conversely, systems which break scaling need the initial conditions to set,
through the initial correlation length, the relative amplitudes
of the multiple growing length-scales.  [The earlier argument
against $\xi_0$ dependence does not apply to systems which break scaling,
since changing $\xi_0$ in such systems is no longer simply equivalent to
an offset in the time-coordinate, but also changes the relative amplitudes
of the growing length scales.] \ It is possible, but unlikely, that multiple
length scales will have relative amplitudes set only through the core scale
$\xi$. So we expect scaling violations if and only if we observe $\xi_0$
dependence in the amplitude of the growth law.  For scaling systems,
therefore, we may set the second argument of the function $F$ in
(\ref{EQN:DIMANAL}) to zero. Then, in cases where no topological defects
are present, or the defects do not dominate the dynamics, $L(t)$ will not
depend on $\xi$ either, and (\ref{EQN:DIMANAL}) gives
$L(t) \sim t^{1/(2+\mu)}$. This is the generic result, covering the region
$n>2$ in figure \ref{FIG:SHORTRANGE}. For $n \le 2$, the defects {\em do}
dominate the dynamics (in that they dominate the energy density, as we
shall see in section \ref{sec:shortrange}), and $L(t)$ can  depend on the
core scale $\xi$. The remainder of the paper is mainly
devoted to understanding these cases.

\section{Growth Laws}
\label{sec:growth}

\subsection{Scaling assumption}
Many phase-ordering systems are empirically found to scale at late times
from a quench, after initial transients have decayed.
Accordingly, the correlation function of the order parameter, $C({\bf r},t) =
\left< \vec{\phi}({\bf x},t)\cdot \vec{\phi}({\bf x}+{\bf r},t) \right>$,
exhibits the scaling form $C({\bf r},t) = f(r/L(t))$,
with a single characteristic length scale, $L(t)$, and a time-independent
scaling function $f(x)$.
This is the dynamic scaling hypothesis \cite{Furukawa85}.
Fourier transforming the scaling form of the correlation function,
we obtain the scaling form of the structure factor,
\begin{equation}
\label{EQN:STRUCT}
S({\bf k},t) = L(t)^d\,g(kL(t)),
\end{equation}
where $d$ is the spatial dimension and $g(y)=\int d^d x e^{i\bf{x} \cdot
\bf{y}} f(x)$.
This scaling hypothesis can be generalized to two-time correlations
by dimensional analysis,
\begin{eqnarray}
\label{EQN:TTSTRUCT}
S({\bf k},t,t') &\equiv&
\left< \vec{\phi}_{\bf k} (t) \cdot \vec{\phi}_{-{\bf k}} (t') \right>,
\nonumber \\
&=& k^{-d} \tilde{g}(kL(t),kL(t'),t/t'),
\end{eqnarray}
so that $g(y) \equiv y^{-d} \tilde{g}(y,y,1)$. From this
we obtain the scaling form of $T({\bf k},t)$,
\begin{eqnarray}
\label{EQN:TWOTIMES}
	\left< \partial_t \vec{\phi}_{\bf k} \cdot \partial_t
	\vec{\phi}_{-{\bf k}} \right>
	&=& \left. \frac{\partial^2}
	{\partial t \partial t'} \right|_{t=t'}
	\left< \vec{\phi}_{\bf k}(t) \cdot \vec{\phi}_{-{\bf k}}(t') \right>,
	\nonumber \\
	&=& \dot{L}^2 k^{2-d} h(kL),
\end{eqnarray}
where $h(x)$ is a new scaling function \cite{ignore}.

Our Energy-Scaling approach is based on the scaling behavior in equations
(\ref{EQN:STRUCT}) and (\ref{EQN:TWOTIMES}). Since the structure of the
system determines the energy, we can obtain the energy as a function of
$L$. The time dependence of the energy will come solely
through $L(t)$, since the structure is invariant up to rescaling. However, the
rate of energy dissipation is  directly determined by $T({\bf k},t)$ and can
be calculated as a function of $L$ and $\dot{L}$,
with no other time dependence.
By equating the time derivative of $\epsilon$ from (\ref{EQN:EN}) with
$\dot{\epsilon}$ from (\ref{EQN:ENDISS}),
we self-consistently obtain $\dot{L}$ as
a function of $L$. From this we solve for the
growth law of the length scale, $L(t)$.
We first consider systems where $n \leq d$, which have
singular topological defects \cite{Toulouse76}.
We will discuss systems without topological defects ($n>d+1$) in
section \ref{sec:defectfree} and
systems with non-singular topological textures ($n=d+1$) in section
\ref{sec:textures}.

\subsection{Energy density}
\label{sec:shortrange}

To calculate the energy-density
we use the scaling form of the structure function (\ref{EQN:STRUCT}).
If the thermodynamic limit of the phase-ordering system exists, then
the infra-red (IR) limit of the integral (\ref{EQN:EN}) is well behaved.
Hence either the momentum integral converges in the ultra-violet (UV)
and $\epsilon \sim L^{-2}$ is extracted using the scaling form and
a change of variables, or momenta on the order of the UV cutoff
dominate the integral.

When structure in the UV limit dominates the integral,
we need to know the behavior of the correlation
function in that limit, with $kL \gg 1$.
This small-scale structure will only come from topological defects,
since small-scale non-defect structures relax quickly
via the dissipative dynamics. As a result, the structure factor is
proportional to the density of defect core,
$S({\bf k},t) \propto \rho_{\rm def}$ for $kL \gg 1$.
Since in an $n$-component model in $d$-dimensions the defect core will
have dimension $d-n$, it
follows from scaling that the core density $\rho_{\rm def} \sim L^{d-n}/L^d
\sim L^{-n}$ \cite{Bray93a}. Hence $S({\bf k},t) \sim L^{-n}$, and the
scaling form (\ref{EQN:STRUCT}) implies
\cite{Bray93a,Bray93g,Bray91a,Toyoki92}
\begin{equation}
\label{EQN:POROD}
S({\bf k},t) \sim L^{-n}\,k^{-(d+n)}\ ,\ \ \ \  kL \gg 1.
\end{equation}
The form of the Porod tail is purely geometrical in origin
and does not depend on the overall defect structure \cite{Bray93g}, or on the
details of the defect core (this holds even for defects with asymmetric
time-independent core structure, see \cite{Rutenberg94c}).
This generalized Porod's law is valid when $n \leq d$,
so that singular topological defects exist.

Using the asymptotic expression (\ref{EQN:POROD}) as needed in the
integral for the energy density (\ref{EQN:EN}), and
imposing a UV cutoff at $k \sim 1/\xi$, we obtain  \cite{Toyoki92}
\begin{eqnarray}
\label{EQN:E}
\epsilon
& \sim & \left\{ \begin{array}{c}
	L^{-n}\,\xi^{n-2}\ ,\ \ \ \ \ \ \ \ \ \ \ \ n<2\ , \\
	L^{-2}\,\ln(L/\xi)\ ,\  \ \ \ \ \ \ \ n=2\ , \\
	L^{-2}\ ,\ \ \ \ \ \ \ \ \ \ \ \ \ \ \ \ \ \ \ n>2\ .
	\end{array} \right.
\end{eqnarray}
The integral (\ref{EQN:EN})
is UV divergent for $n \leq 2$ and convergent for $n>2$.
For $n=2$, when the integral is logarithmically divergent, we impose a
effective lower
cutoff at $k \sim 1/L$, which is the length scale at which (\ref{EQN:POROD})
breaks down and which reflects
the relative lack of structure at scales longer than $L(t)$.
We see that the energy is dominated by the defect core density,
$\rho_{\rm def}$, for $n<2$, by the defect field at
all length scales for $n=2$,
and by variations of the order parameter at scale $L(t)$ for $n>2$.

\subsection{Energy-density dissipation}
\label{sec:ediss}
We calculate the scaling  behavior of the rate of energy-density dissipation,
$\dot{\epsilon}$, as a function of $L(t)$ in an analogous manner.
If the energy-dissipation integral (\ref{EQN:ENDISS}) converges we use
the scaling form (\ref{EQN:TWOTIMES}) and change variables
to obtain $\dot{\epsilon} \sim L^{\mu-2} \dot{L}^2$.
For the cases when the momentum integral diverges
in the UV, we must evaluate the time-derivative correlation function
in the $kL \gg 1$ limit.

In the UV limit, structure comes from defect cores.
The field sufficiently close to a defect core comoves with the core. If the
core has a local velocity ${\bf v}$, then
\begin{equation}
\label{EQN:COMOVING}
\partial_t{\vec{\phi}} \simeq R_{{\bf \omega}} \vec{\phi} -
	\bf{v}\cdot \nabla \vec{\phi},
\end{equation}
close to the core (i.e.\ at distances small compared to L),
where $R_{{\bf \omega}} $ is a rotation matrix.
In the momentum representation, the second term of (\ref{EQN:COMOVING})
probes the structure of the defect and scales as
$k \dot{L}$, while the rotation term does not and only scales as
$\dot{L}/L$. Hence the rotation term is negligible when $kL \gg 1$.
This gives $\partial_t{\vec{\phi}}_{\bf k} \sim  k v \vec{\phi}_{\bf k}$ in
that limit, and
\begin{eqnarray}
\label{EQN:TTLARGEK}
\left< \partial_t \vec{\phi}_{\bf k} \cdot
	\partial_t \vec{\phi}_{\bf -k}\right> &\sim&
\left< v^2 \right>_k k^2 \left< \vec{\phi}_{\bf k} \cdot
\vec{\phi}_{\bf -k}\right>, \ \ \ \ \ \ \ \ \ kL \gg 1, \nonumber \\
&\sim&
\left< v^2 \right>_k k^{2-d-n} L^{-n}, \ \ \ \ \ \ \ \ \ \ \ \ kL \gg 1,
\end{eqnarray}
where $\left< v^2 \right>_k$ is the square velocity of
defect core, averaged over the core elements that contribute to structure
at momentum $k$.  To obtain the second line we have used the
generalized Porod's law (\ref{EQN:POROD}) and have
implicitly assumed that all lengths are large with respect the core
size: $k^{-1},L \gg \xi$.
We emphasize that (\ref{EQN:COMOVING}) is valid  sufficiently close
to a slowly moving defect core. Since $\partial_t \vec{\phi}$ is purely
dissipative from (\ref{EQN:DYNAMICS}), the defect core in
(\ref{EQN:COMOVING}) must be moving
``downhill'' in the energy landscape of the local order parameter field and,
for extended defects, under the influence of the surface or line-tensions
of the defect core itself.

The characteristic value of the defect velocity $v$ is $\dot{L}$.
If the energy-dissipation is dominated by the evolution of large defects
at the characteristic size $L$, then using Eq. (\ref{EQN:TTLARGEK}) and
replacing $\left< v^2 \right>_k$ with $\dot{L}^2$
within the energy-dissipation integral (\ref{EQN:ENDISS})
is appropriate.
We show below that this is the case for
$n<d$ or for $n>2$. If the integral converges
in the UV this is equivalent to using the scaling form (\ref{EQN:TWOTIMES}).
If the integral diverges in the UV, this implies that the energy dissipation
is due to the slow evolution at characteristic
scales (such as shrinking domains or vortex loops),
leading to a reduction of the core volume. The actual asymptotics
of (\ref{EQN:TTLARGEK}) is in general {\em not} given by using
$\left< v^2 \right>_k \sim \dot{L}^2$, as we will see explicitly in
section \ref{sec:microscopic}, so we must
directly check the appropriateness of the
substitution in calculating $\dot{\epsilon}$.

For $n \leq 2$, when the energetics is dominated by the defects (following
Eq. (\ref{EQN:E})),
the rate of energy dissipation can be written as the rate
of change of the energy of all the defect features \cite{Bray94}:
\begin{eqnarray}
\label{EQN:DISSLENGTH}
   \dot{\epsilon}  &\sim& \frac{\partial}{\partial t}
	\int_\xi^\infty dl\, n(l,t)\, E(l)\,, \nonumber \\
   & =& - \int_\xi^\infty dl \frac{\partial j(l,t)}{ \partial l} \,
	E(l)\,, \nonumber \\
   &= & j(\xi) E(\xi) + \int_\xi^\infty dl\, j(l,t)\,
	\frac{\partial E(l)}{\partial l}\,,
\end{eqnarray}
where $n(l,t)$ is the number density of defect features of scale $l$
\cite{feature}, $E(l)$ is the average energy of a defect feature of scale $l$,
and  $j(l)$ is the number flux of defect
features. We use the continuity equation, $\partial n/ \partial t
+ \partial j/ \partial l =0$, to obtain the second line in
(\ref{EQN:DISSLENGTH}). The total number density of defect features, $N$,
scales as the inverse scale volume, $N \sim 1/L^d$, and hence
$\dot{N}$ is slowly varying for times
of order $L/\dot{L}$. Since defects only vanish at the  core scale $\xi$,
we have $\dot{N} \sim j(\xi)$.
This implies that $j(l)$ is independent of $l$ for
$l \ll L$, in order to provide a steady rate of defect extinction.
We would like to know whether the integral in Eq. (\ref{EQN:DISSLENGTH})
is dominated by small scales, with $l \ll L$, and so we need to know $E(l)$
in that limit.  We assume here that small defect features have an average
energy related to the scaling form (\ref{EQN:E}) for the energy-density:
\begin{equation}
\label{EQN:ENFEATURE}
	E(l) \sim \left\{ \begin{array}{c}
			l^{d-n}, \ \ \ \ \ \ \ \ \ n<2, \\
			l^{d-2} \ln(l/\xi), \ \ \ \  n=2, \\
			l^{d-2}, \ \ \ \ \ \ \ \ \ \ n>2.
		\end{array} \right.
\end{equation}
This is explicitly confirmed for isolated symmetric defects in section
\ref{sec:microscopic}. Using (\ref{EQN:ENFEATURE}), we see that for
$d>n$ or $n>2$ the integral  in (\ref{EQN:DISSLENGTH})
is well behaved at $l \ll L$ and the integral dominates the
$j(\xi)\, E(\xi)$ term. For these cases,
structures with scales and separations
$l \sim L(t)$ always dominate the energy dissipation, so that
$\left< v^2 \right>_k \sim \dot{L}^2$ can be used in evaluating
$\dot{\epsilon}$ with equation (\ref{EQN:TTLARGEK}).
For $d=n < 2$ the integral in (\ref{EQN:DISSLENGTH}) diverges for
$l \ll L$, and dissipation is dominated by defect pairs annihilating.
We treat the case $d=n=1$ below.
For $d=n=2$ the integral is logarithmically divergent, and $\dot{\epsilon}$
has contributions from defect pairs at all scales. We treat this case
in section \ref{sec:microscopic} by determining $\langle v^2 \rangle_k$
directly.

For systems with $n<d$ or $n>2$ we use $\left< v^2 \right>_k \sim \dot{L}^2$
in (\ref{EQN:TTLARGEK}), {\em within} the energy-dissipation integral
(\ref{EQN:ENDISS}). This gives
\begin{eqnarray}
\label{EQN:RHS}
\dot{\epsilon}
& \sim &  \left\{ \begin{array}{c}
	L^{-n}\,\xi^{n+\mu-2}\, \dot{L}^2\ ,\ \ \ \ n+\mu<2\ , \\
	L^{-n}\,\ln (L/\xi)\, \dot{L}^2\ ,\ \ \ n+\mu=2\ , \\
 	L^{\mu-2}\, \dot{L}^2\ ,\ \ \ \ \ \ \ \ \ \ \ \ \ n+\mu>2\ ,
	\end{array} \right.
\end{eqnarray}
apart from $n=d \leq 2$.
For $n+\mu < 2$ the integral is UV divergent and we have
used the $kL \gg 1$ form (\ref{EQN:TTLARGEK})
between its lower limit of applicability, $k \gtrsim
L^{-1}$, and the UV cutoff at $k \sim 1/\xi$. For $n+\mu =2 $ the
integral is logarithmically divergent. Again we use the $kL \gg 1$ asymptotics
in its region of applicability. [ The scaling contribution from $k L \sim 1$
merely changes the effective $\xi$ in the logarithm of (\ref{EQN:RHS}).]
For $n+\mu > 2$ the integral is
convergent and we use the scaling form (\ref{EQN:TWOTIMES}).
Equating $\dot{\epsilon}$ from (\ref{EQN:RHS}) to the time derivative
of $\epsilon$ from (\ref{EQN:E}), we obtain the growth laws shown in
figure \ref{FIG:SHORTRANGE}.

As an example, consider nonconserved scalar fields. For this case, with
$\mu=0$ and  $n=1$, Eq.\ (\ref{EQN:E}) gives $\epsilon \sim 1/L\xi$,
implying $\dot{\epsilon} \sim -\dot{L}/L^2\xi$, while (\ref{EQN:RHS})
gives $\dot{\epsilon} \sim -\dot{L}^2/L\xi$. Equating these two results
for $\dot{\epsilon}$ yields $\dot{L} \sim 1/L$, implying $L \sim t^{1/2}$.
The other results displayed in figure \ref{FIG:SHORTRANGE} were obtained
in the same way.

For 1D scalar systems ($n=d=1$), equal-time correlations scale
\cite{Rutenberg94a,Nagai83} but $T({\bf k},t)$ {\em breaks} scaling
due to rapidly annihilating domain wall pairs \cite{Rutenberg94a}.
We can still apply the Energy-Scaling argument to the exponentially
suppressed interaction energy between domain wall pairs, expressed
as an energy density, $\epsilon_{\text{int}} \sim e^{-L/\xi}/L$. We
equate the time-derivative of
this to the energy-dissipation from equation (\ref{EQN:RHS})
(since the interaction energy changes from the slow $\left<v^2 \right> \sim
\dot{L}^2$ evolution of the defects). This leads to logarithmic growth
$L(t) \sim \ln{t}$ for both non-conserved and conserved scalar models in
one-dimension, as shown in table \ref{TAB:2DXYGROWTH}.

The growth laws shown in figure \ref{FIG:SHORTRANGE} and table
\ref{TAB:2DXYGROWTH} are only the leading time-dependence
of the properly integrated equations of motion for
the length scale. Further corrections come from subdominant parts of the
energy integrals, and from subdominant corrections to the scaling forms
of the correlation functions. Corrections to integrals dominated
by the UV limit are given by contributions from $kL \sim 1$,
and by corrections
to the generalized Porod's law (\ref{EQN:POROD}). The latter are not
generally known, except for three-dimensional scalar systems
where the leading correction to
Porod's law is $L^{-3} k^{-6}$ \cite{Tomita84}.
Corrections to convergent or logarithmically divergent integrals can come
from the UV cutoff. For instance, for non-conserved XY systems the $\ln L$
factors in (\ref{EQN:E}) and (\ref{EQN:RHS}) will in general have
different effective cutoffs, of order the core size. This leads to an
additive correction to the growth law, $O(L(t)/\ln{t})$.
It is important to be aware of such corrections for each particular system.

\subsection{Long-range attractive interactions}
\label{sec:long-range}

For systems with long-range {\em attractive} interactions
\cite{Rutenberg94a,Hayakawa93a,Lee93,Bray93b}
the rate of energy-density dissipation will still be given by
(\ref{EQN:ENDISS}), but the energy density has a new contribution
\begin{equation}
\label{EQN:LONGRANGE}
\epsilon_{\text{LR}} \sim \int_{\bf k} k^\sigma\,
\left< \vec{\phi}_{\bf k} \cdot \vec{\phi}_{-{\bf k}} \right>,
\end{equation}
with $0 < \sigma \leq 2$. This
reduces to the short-range case for $\sigma=2$, and is subdominant
to the short-range interactions for $\sigma>2$.  This interaction can be
motivated by a term in the energy-functional:
\begin{equation}
\label{EQN:INTERACTION}
	H_{\text{LR}} = \int d^dx \int d^dr
	\frac{\left[
	\vec{\phi}({\bf x}+{\bf r})-\vec{\phi}({\bf x})\right]^2}
		{ r^{d+\sigma}},
\end{equation}
where we take $\sigma >0$ for a well defined thermodynamic limit.
As in the short-range case, the
local potential $V(\vec{\phi})$ does not dominate the
gradient, and can be neglected. We see this for $0 < \sigma \leq 2$ by
generalizing equation (\ref{EQN:AMPEN})
to balance a long-range energy $(1-\delta)^2 L^{-\sigma}$ with the potential
term $4 A \delta^2$. The local energy density is minimized if
$\delta \sim 1/L^\sigma$, where $|\vec{\phi}| = 1-\delta$. This
corresponds to a $1/r^\sigma$ tail in the profile of an
isolated defect, a generalization of the $1/r^2$ tail for vector
systems with only short-range
interactions. This tail does not change the generalized Porod's law
(\ref{EQN:POROD}) for either short- or long-range interactions,
because the local structure
of the defect does not change as the length scale $L(t)$ grows. As before,
the structure factor for  $k L \gg 1$ is simply proportional to the density
of defect core. Using Porod's law in (\ref{EQN:LONGRANGE}), we find
\begin{eqnarray}
\label{EQN:ELR}
\epsilon_{\text{LR}} & \sim & \left\{ \begin{array}{c}
	L^{-n}\,\xi^{n-\sigma}\ ,\ \ \ \ \ \ \ \ \ \ \ \ n<\sigma, \\
	L^{-\sigma}\,\ln(L/\xi)\ ,\ \ \ \ \ \ \ \ n=\sigma,\\
	L^{-\sigma}\ ,\ \ \ \ \ \ \ \ \ \ \ \ \ \ \ \ \ \ \  n>\sigma.
	\end{array} \right.
\end{eqnarray}
Comparing with the energy-density from the short-range interactions
(\ref{EQN:E}), we see that $\epsilon_{\text{LR}}$ always
describes the scaling of the energy density at late times
for $0<\sigma \leq 2$.

The asymptotics of $T({\bf k},t)$ are still given by (\ref{EQN:TTLARGEK}).
Applying Eq.\ (\ref{EQN:DISSLENGTH}) to the long-range case shows that
we can use $\left<v^2 \right>_k \sim \dot{L}^2$ {\em within} the
energy-dissipation integral (\ref{EQN:ENDISS}) for all cases except
$n=d \le \sigma$ \cite{Bray94}. Comparing the rate of energy dissipation,
still given by (\ref{EQN:RHS}), but excluding $n=d \leq \sigma$, and the
time derivative of (\ref{EQN:ELR}), we find $\dot{L}$ and hence $L(t)$.
The results are summarized in figure \ref{FIG:LONGRANGE}. We treat the cases
$n=d=1 \leq \sigma <2$,
where $\left< v^2 \right>_k$ is dominated by small defect features,
elsewhere and find $L(t) \sim t^{1/(1+\sigma)}$ \cite{Rutenberg94a}.

\section{Special Cases}
\label{sec:special}

\subsection{Systems without Topological Defects}
\label{sec:defectfree}

Systems with no topological defects [such as $O(n)$ models with $n>d+1$]
will have no power-law UV structure, and hence will have convergent energy
integrals, giving $L(t) \sim t^{1/(\sigma+\mu)}$, where $\sigma=2$ for
short-range interactions. This result for $L(t)$ also follows directly from
the dimensional analysis discussed at the end of section \ref{sec:model},
if the system scales.
Any power-law UV structure would indicate a scaling density of singular
structure --- which we do not expect if locally stable defects are absent.
For instance, Rao and Chakrabarti find stretched
exponential tails to the asymptotic structure factor \cite{Rao94} for 2D $n=4$
systems.  Of course, this result applies to any system with initial
conditions that evolve into a scaling structure without topological
defects, and not just to $n>d+1$. For example, if the 2D XY model is quenched
{\em from} below $T_{KT}$ then the bound vortex pairs quickly annihilate and
the scaling state has no topological defects (for a quench to $T=0$)
\cite{Rutenberg94d}. In the non-conserved case scaling is obeyed and the growth
law is $L \sim t^{1/2}$ \cite{Rutenberg94d}, as expected.  In the conserved
2D XY model quenched from below $T_{KT}$, $t^{1/4}$ growth (or, more
generally, $t^{1/(\sigma+\mu)}$) will obtain, provided scaling holds.
Of course, scaling can be violated even without topological defects --- as
seen in the conserved spherical model \cite{Coniglio89}.

\subsection{Isolated Defects}
\label{sec:microscopic}

In this section we obtain the size $l(\tilde{t})$, as a function of time
to collapse $\tilde{t}$, of an isolated symmetric
defect. We will then use this, for scalar and 2D XY systems, to
obtain the proper asymptotics of $T(k,t)$, i.e. an expression for
$\langle v^2 \rangle_k$ in (\ref{EQN:TTLARGEK}). For scalar systems this
will confirm the growth laws we have obtained. For 2D XY systems, we
will be able to obtain the scaling growth laws
(see table \ref{TAB:2DXYGROWTH}) where previously we
were prevented by our crude knowledge of $\langle v^2 \rangle_k$. [In the
following section we will apply a similar approach to systems with
non-singular topological textures.]

Applying equations (\ref{EQN:EN}) and (\ref{EQN:ENDISS}) to
isolated defects, in a ``microscopic'' approach, is appealing but
not trivial.  For scaling systems, with many defects,
the disordered initial conditions will cut off
any momentum integral in the infra-red (IR).  Equivalently, there is no
structure at scales much greater than the inverse
of the characteristic length, $L^{-1}$, which thus provides a natural
IR cutoff.  On the other hand, for solitary defects (such as a single
spherical domain), an IR cutoff is provided by the
structure and evolution of the single defect itself.
We will apply the quasi-adiabatic
assumption to the defect structure: that the order-parameter
field is locally equilibrated to the defect core configuration. The breakdown
of this assumption away from the core, due to a finite core velocity, will
provide a natural IR cutoff --- cutting off the structure at small $k$,
corresponding to large spatial scales.
We must apply the largest appropriate IR cutoff, and we will focus
on small defects within larger phase-ordering systems, where both cutoffs
to the IR behavior are appropriate. Whenever possible, we check our
approach with established results.  We restrict our attention to the
collapse of a spherical scalar domain of radius $l$ with $d>1$,
and the annihilation of a point
defect/anti-defect pair of separation $l$ in a 2D XY model ($n=d=2$).
We calculate both the structure factor $S({\bf k},l)$ and the time-derivative
correlations $T({\bf k},l)$, use these in the independent
energy equations (\ref{EQN:EN}) and (\ref{EQN:ENDISS}), and self-consistently
solve for $l(\tilde{t})$ \cite{Siggia81}.

We first consider generic systems with an isolated, symmetric, singular
topological defect structure ($n \leq d$) \cite{Toulouse76}
characterized by a single length-scale much
larger than the core size, $l \gg \xi$.  For example, spherical
domain walls, or a point defect/anti-defect pair.
The defect size, $l$, corresponds to the local radius of curvature
for $n<d$ or to the separation of point-defects for $n=d$.
The structure-factor at small scales,  $k l \gg 1$,
is proportional to the core volume of the defect, $l^{d-n}$, and has the
asymptotic momentum dependence $k^{-(d+n)}$ \cite{Bray93g}
of a ``flat'' or stable defect structure.
[This asymptotic momentum dependence
leads to a generalized Porod's law for scaling systems
\cite{Bray93a,Bray93g,Bray91a,Toyoki92}.]
For $k l \gg 1$ the field around an element of defect core will
comove with it [without a rotation term because of the symmetry of the defect]:
$\partial_t \vec{\phi} = (dl/dt)(\nabla \vec{\phi})_{\text{radial}}$,
where we have taken the radial component of the gradient.
We then have, for the isolated defect,
\begin{eqnarray}
\label{EQN:UV}
	S({\bf k},l)
		& = & l^{d-n}\,k^{-(d+n)}\,f_S(kl)\ ,
\nonumber \\
	T({\bf k},l)
		& = & (dl/dt)^2\,l^{d-n}\,k^{-(d+n-2)}\,f_T(kl)\ ,
\end{eqnarray}
where the scaling functions $f_S(x)$ and $f_T(x)$ tend to constants for
large $x$.  From (\ref{EQN:EN}), the
energy of the isolated defect of scale $l \gg \xi$ is given by
\begin{eqnarray}
\label{EQN:MEN}
 E(l) &\sim& \int_{\bf k} k^2 S({\bf k},l),  \nonumber \\
   &\sim& \int_0^{\xi^{-1}} d^d k\,l^{d-n}\,k^{-(d+n-2)}\,f_S(kl)\ ,
\end{eqnarray}
where we cut off the integral above at the inverse core size.
The energy integral (\ref{EQN:MEN}) always converges at small $k$, as will be
checked on a case by case basis.  Evaluating the integral gives
\begin{eqnarray}
\label{EQN:MENEVAL}
E(l) &\sim & \left\{ \begin{array}{c}
	l^{d-1} \xi^{-1}, \ \ \ \ \ \ \ \ \ \ \ \ \ \ n=1, \\
	l^{d-2} \ln{(l/\xi)}, \ \ \ \ \ \ \ \ \ \ n=2, \\
	l^{d-2}, \ \ \ \ \ \ \  \ \ \ \ \ \ \ n > 2 .
	\end{array} \right.
\end{eqnarray}
[We include $n>2$ for completeness. Isolated defects with $n>2$
will generically be asymmetric \cite{Rutenberg94c,Ostlund81}, and not
described by a single scale $l$.] The results for $n=1,2$ are, of course,
familiar: for $n=1$ the energy is just proportional to the surface area
$l^{d-1}$ of the domain, while the factor $\xi^{-1}$ is an estimate of the
surface tension; for $n=2$, the energy is given by the defect core volume
$l^{d-2}$ multiplied by a logarithmic factor from the `far field', with
a short-distance cut-off at the core scale $\xi$ and a large-distance cut-off
at the size $l$ of the defect, reflecting the self-screening
of the defect (a vortex loop for $d=3$, or vortex-antivortex pair for $d=2$)
at large distances.

Next we evaluate the analogue of equation (\ref{EQN:ENDISS}), the energy
dissipation rate:
\begin{eqnarray}
\label{EQN:MENDISS}
\dot{E}(l) & = & -\int_0^{\xi^{-1}} d^dk\,k^{-\mu}\,T({\bf k},l) \nonumber \\
	   &\simeq &
- \int_{k_{\text{cut}}}^{l^{-1}} d^d k\, k^{-\mu}\, T({\bf k},l)
- \int_{l^{-1}}^{\xi^{-1}} d^d k\, k^{-\mu}\, T({\bf k},l) \nonumber \\
&\equiv& 	\dot{E}_{IR}(l) + \dot{E}_{UV}(l)\ ,
\end{eqnarray}
where $\dot{E}_{IR}$ and $\dot{E}_{UV}$ correspond to the momentum
integrals over $k < l^{-1}$ and $k > l^{-1}$ respectively.  The $kl \gg 1$
behavior, given by
equation (\ref{EQN:UV}) with $f_T(kl)=const.$, can be used to
calculate $\dot{E}_{UV}$.
However, we will see that, unlike in the calculation of $E(l)$,
the IR behavior (through $\dot{E}_{IR}$) sometimes dominates
$\dot{E}(l)$, and so we need to determine both the IR structure of
$T({\bf k},l)$  and the effective IR cutoff, $k_{cut}$.
For scalar systems, $\dot{E}_{IR}$ is only
significant for conservation laws with $\mu \geq d$, and
$k_{\text{cut}}$ depends on whether or not the isolated domain is in
a larger phase-ordering system or not.  For 2D XY systems,
$\dot{E}_{IR}$ is always significant, and $k_{\text{cut}}$ is
provided by the breakdown of the quasi-adiabatic approximation at large
distances due to the non-zero evolution velocity of the defect core.

\subsubsection{Scalar Systems}

In scalar systems with $d>1$, we consider a spherical domain with radius $l$.
Under the quasi-adiabatic approximation, the field inside and outside
the domain are taken to minimize the energy subject to the configuration
of the domain wall. In the IR limit, with $kl \ll 1$, the field is dominated
by the saturated value inside and outside the domain walls and so
$\phi_{k} \sim l^{d}$. For conserved fields we know that
$\phi_{k=0}$ does {\em not} evolve with $l$ and so the quasi-adiabatic
approximation must break down as $k \rightarrow 0$.  The finite evolution
rate of the domain will determine the effective $k_{cut}$.
Using $\phi_k \sim l^d$, we easily obtain the IR asymptotics:
\begin{eqnarray}
\label{EQN:SCALARIR}
	S({\bf k},l) &\sim& l^{2d}, \ \ \ \
	\ \ \ \ \ \ \ \ \ \ \ \ \ \ \ \ \ \ \ k_{cut} < k \ll l^{-1},
\nonumber \\
	T({\bf k},l) &\sim& (dl/dt)^2 l^{2d-2},
		\ \ \ \ \ \ \ \ \ \ \ \ k_{cut} < k \ll l^{-1},
\end{eqnarray}
where we have not yet determined $k_{cut}$.  This IR limit, combined with the
UV
behavior from (\ref{EQN:UV}) and the appropriate $k_{cut}$, will be sufficient
to determine the evolution of an isolated domain. Note that the small $kl$
forms (\ref{EQN:SCALARIR}) are consistent with the general scaling forms
(\ref{EQN:UV}).
In principle the complete scaling functions, $f_S$ and $f_T$,
can be calculated. In practice,
however, we do not need the precise structure for $k l \sim 1$, as its scaling
properties with $l$ are embodied in the IR and UV limits.

For non-conserved systems, the IR asymptotics imply that
the energy-dissipation integral (\ref{EQN:MENDISS}) converges
in the IR and is dominated by $\dot{E}_{UV}$, leading to
\begin{equation}
\label{EQN:SCALEDISEVAL}
	\dot{E} \sim - (dl/dt)^2 l^{d-1} \xi^{-1}.
\end{equation}
Comparing this with the time derivative of (\ref{EQN:MENEVAL}),
recovers the standard result \cite{Allen79} $dl/dt \sim - l^{-1}$
for $d>1$, which gives
\begin{eqnarray}
\label{EQN:SCALECOLLAPSE}
	l(\tilde{t}) &\sim &
		\tilde{t}^{1/2},
\end{eqnarray}
where $\tilde{t} = t_{\text{max}}-t$ is the time left before
the defect annihilates.

We can also extract a scale-dependent effective
mobility for non-conserved systems \cite{mobility}, $\eta(l)$, by taking
\begin{equation}
\label{EQN:MOBILITY}
	\frac{dl}{dt}\sim - \eta \frac{\partial E/\partial l}{l ^{d-n}},
\end{equation}
so that $\eta(l)$ is the proportionality factor between the driving force
per core volume and the evolution speed. This gives $\eta \sim \xi$ for
$n=1<d$. While the mobility of a
scalar system depends on the details of the potential through $\xi$, i.e.
on the surface tension, the evolution velocity $dl/dt$ is independent of those
details (for free energy functionals of the form (\ref{EQN:HAMILTONIAN}));
this was first observed by Allen and Cahn \cite{Allen79}.

For conserved systems with $\mu <d$, the IR limit still does not dominate
$\dot{E}$ and the dependence on $l$ is still captured by $\dot{E}_{UV}$.
The results are summarized in table \ref{TAB:SCALAR}:  for $\mu<1$ the UV
limit dominates the energy-dissipation integral; for $\mu =1$ the integral
is logarithmically divergent and is cut off by the detailed
structure at $ k l \sim 1$; for $\mu >1$ the integral of $\dot{E}$
converges in the UV, and structure at $k l \sim 1$
dominates. The scaling behavior of this contribution is given by power
counting on the second line of (\ref{EQN:MENDISS}).

For strong-enough conservation laws, $\mu \geq d$, the
energy-dissipation integral no longer converges at small $k$. It
is dominated by the IR behavior of $\dot{E}_{IR}$ and we need to know the
appropriate cutoff $k_{\text{cut}}$.  First we consider the breakdown of the
quasi-adiabatic approximation far from the defect. This provides the
only cutoff for a solitary domain in an infinite system. The cutoff is
provided by the behavior of the field $\phi$ far from the defect, where
we linearize the evolution equation around the ordered state
$|\vec{\phi}|=1$ and can neglect all but the lowest derivatives.
{}From equation (\ref{EQN:DYNAMICS}) we have
\begin{equation}
\label{EQN:LINEAR}
	\partial_t (\delta \phi_{\bf k}) = -8 A k^\mu \delta
		\phi_{\bf k}, \ \ \ \ \ kl \ll 1.
\end{equation}
Any breakdown of quasi-adiabaticity (i.e. the field not following the
static configuration given by the configuration of the domain wall) is
due to the finite evolution speed $dl/dt$.
Dimensionally from Eq. (\ref{EQN:LINEAR}),
$dl/dt \sim A k_{\text{vel}}^{\mu-1}$,
which determines a velocity-dependent IR cutoff $k_{\text{vel}} \sim
(dl/dt)^{1/(\mu-1)}$.   This cutoff can be used for $k_{cut}$ in
Eq. (\ref{EQN:MENDISS}) to determine the evolution of an isolated domain
(with $l \gg \xi$) in an infinite system.  However, within the context
of a larger phase-ordering system, other evolving domains provide
a natural cutoff $k_{\text{cut}} \sim L^{-1}$, where $L$ is the
characteristic length-scale of the system.  This cutoff can be used to
determine the $l$-dependence of $dl/dt$ (i.e. the ``scaling'' behavior of the
domain), again with Eq. (\ref{EQN:MENDISS}). The two results can be compared
and we find that the velocity cutoff is only larger than $L^{-1}$ for
domains with $l \lesssim L^{1-1/d}$.
This is a vanishing proportion of domains in the
scaling limit, so we have used $k_{\text{cut}} \sim L^{-1}$ in table
\ref{TAB:SCALAR}. [The difference between the cutoffs does imply that
for scalar systems with
$\mu \geq d$ a solitary domain will evolve qualitatively differently
than a small domain in a phase-ordering system.]
Comparing the dominant energy-dissipation
($\dot{E}_{IR}$ in this case) with the time-derivative of
(\ref{EQN:MENEVAL}), we find table \ref{TAB:SCALAR} for $d>1$.
We obtain the standard $l \sim \tilde{t}^{1/3}$ for $\mu=2<d$ \cite{Langer92},
and recover the $\mu=d=2$ result of Rogers and Desai \cite{Rogers89} for small
domains: $dl/dt \sim -1/[l^2 \ln{(L/l)}]$, where, unlike ref. \cite{Rogers89},
we have retained the defect scale in the logarithm.

We can estimate $\left< v^2 \right>_k$ for scalar systems with our
results in Table \ref{TAB:SCALAR}. This will enable us to determine the
UV asymptotics of $T(k,t)$ directly through (\ref{EQN:TTLARGEK}), and with that
to check our previous results and the consistency of our approach.
We assume that small
defect features, with $l \ll L$, evolve as described in Table \ref{TAB:SCALAR}
with the local rms velocity of the defect core given by $v(l) \sim dl/dt$.
We further assume that small defect features will evolve, at least in their
$l$ dependence, like segments of
spherical domains, with a local radius of curvature decreasing as
$dl/dt$, so that the number flux is given by $j(l)=n(l)v(l)$. Then
\begin{eqnarray}
\label{EQN:VELSQUARE}
\left< v^2 \right>_k &\simeq& \frac{\int_{l_{\text{min}} }^L dl \,
		v^2(l) n(l) l^{d-n}}
		{\int dl \, n(l) l^{d-n}}, \nonumber \\
	 &\sim & \dot{L} L^{n-d-1} \int_{l_{\text{min}}}^L dl \,
	| v(l) | l^{d-n},
	\ \ \ \ \ \ \ \ \ \ \ kL \gg 1.
\end{eqnarray}
If small features ($l \ll L$) dominate we use
$ j(l) \sim j(\xi) \sim - \dot{L} L^{-(d+1)}$ in (\ref{EQN:VELSQUARE}),
while if small features do not dominate then the same approximation will give
the expected $\langle v^2 \rangle_k \sim \dot{L}^2$.
To evaluate the integral, with $n=1$ for scalar systems,
we take $v(l) \sim dl/dt$ for $l \ll L$ and use table \ref{TAB:SCALAR}.
Only features large enough to
contribute to the asymptotic structure of $T(k,t)$ are counted in
$\langle v^2 \rangle_k$, and this sets $l_{min}$ in (\ref{EQN:VELSQUARE}).
Comparing Eq. (\ref{EQN:TTLARGEK}) and (\ref{EQN:UV}) we
see that $l_{\text{min}} = k^{-1}$.  We use this
in equation (\ref{EQN:VELSQUARE}) and
summarize our results in table \ref{TAB:SRVELSQUARE}. These results,
used in equation (\ref{EQN:TTLARGEK}), give
the true $kL \gg 1$ limit of $T({\bf k},t)$.  These asymptotics are
{\em not} given by simply approximating $\langle v^2 \rangle_k \sim \dot{L}^2$,
at least for $\mu \geq d$.
However, the true asymptotics, used in (\ref{EQN:ENDISS}),
give the same growth laws as obtained earlier with our less detailed
approach in section \ref{sec:growth}
(which has the advantage of requiring fewer assumptions about the dynamics of
small defect features within the phase-ordering system).
For $\mu \leq 1$, where $\dot{\epsilon}$ is dominated by the UV asymptotics
of $T(k,t)$, then the proper asymptotics show that
$\langle v^2 \rangle_k \sim \dot{L}^2$.  Conversely, for
$\mu \geq d$, where $\langle v^2 \rangle_k$ is {\em not} given by $\dot{L}^2$,
we confirm that $\dot{\epsilon}$ is not controlled by the UV asymptotics
of $T(k,t)$. Hence, the details of our treatment in section
\ref{sec:growth}, as well as the growth laws, are consistent with this more
microscopic treatment.

\subsubsection{XY Systems}

We can apply the same quasi-adiabatic assumption to the 2D XY model: that
the order-parameter field is given by the static configuration with boundary
conditions imposed by the defect configuration.
For the 2D XY model, the hard-spin ($V_0 \rightarrow \infty$) energy is
optimized by XY phases which satisfy $\nabla^2 \theta =0$ away from the
vortices.  For an isolated defect/anti-defect pair separated
by $l$, this leads to
\begin{equation}
\label{EQN:2DXYFIELD}
	\theta({\bf r}) = \tan^{-1} \left[ \frac{y}{x-l/2} \right]
		- \tan^{-1} \left[ \frac{y}{x+l/2} \right],
\end{equation}
where ${\bf r}=(x,y)$ and the order-parameter
$\vec{\phi}({\bf r})= (\cos{\theta},\sin{\theta})$ \cite{metastable}.
For quasi-adiabatic evolution, the time dependence is only
through $l$, and the IR behavior of the spin-spin correlations is
straight-forward to determine from (\ref{EQN:2DXYFIELD}):
\begin{eqnarray}
\label{EQN:2DXYIR}
	S({\bf k},l) &\sim& l^2/k^2, \ \ \ \ \ \
		\ \ \ \ \ \ \ \ \ \ kl \ll 1, \nonumber \\
	T({\bf k},l) &\sim& (dl/dt)^2/k^2, \ \ \ \ \ \ \ kl \ll 1.
\end{eqnarray}
Note once more that these results are consistent with the general scaling
forms (\ref{EQN:UV}). Using (\ref{EQN:2DXYIR}), the energy integral converges
in the IR and is given by (\ref{EQN:MENEVAL}).
For non-conserved systems, both $\dot{E}_{IR}$ and $\dot{E}_{UV}$
(\ref{EQN:MENDISS}) are logarithmically divergent. For conserved
systems $\dot{E}_{IR}$ dominates and is infra-red divergent. For both
cases we must determine the appropriate IR cutoff $k_{\text{cut}}$.

The field far from the vortices can only
can only follow their motion for infinitesimal $dl/dt$.
For non-zero evolution speeds, the failure of the quasi-adiabatic
approximation far from the vortices provides an infra-red cutoff.
Far from the vortices, the component of $\partial_t \vec{\phi}$
parallel to the field is still dominated
by Eq. (\ref{EQN:LINEAR}), but the transverse component is given
by the gradient term:
$(\partial_t \vec{\phi})_{\perp} \sim -(-\nabla^2)^{\mu/2} \nabla^2
\vec{\phi}$. Dimensionally from this equation,
the transverse component provides a
larger IR cutoff than Eq. (\ref{EQN:LINEAR}),  so that
$dl/dt \sim k_{\text{cut}}^{\mu+1}$.  This breakdown of
quasi-adiabaticity implies an IR cutoff
$k_{\text{cut}} \sim (dl/dt)^{1/(1+\mu)}$. This cutoff can be used
with the IR asymptotics (\ref{EQN:2DXYIR}) in the energy-dissipation
integrals of Eq. (\ref{EQN:MENDISS}) to determine the evolution of a
solitary vortex pair. We summarize the results in table \ref{TAB:2DXYMICRO}.
For closely separated vortex pairs within a phase-ordering
system, other defect pairs will provide an effective cutoff at $L^{-1}$.
The largest cutoff will apply in Eq. (\ref{EQN:MENDISS}).   We find that for
non-conserved systems, the breakdown of quasi-adiabaticity provides
the dominant cutoff for $l \lesssim L / \ln{(L/\xi)}$,
and for conserved systems for $l \lesssim L$.  Since in the non-conserved
case $k_{\text{cut}}$ only enters logarithmically, for conserved and
non-conserved systems any defect with $l \ll L$ will have an
effective IR cutoff provided by the breakdown of quasi-adiabaticity.
For non-conserved systems, since the cutoff only enters logarithmically,
the velocity screening is effectively the same as the length-scale
screening used by Yurke {\em et al} \cite{Yurke93}, though from
(\ref{EQN:2DXYIR}) we see that $T({\bf k},l)$ for the defect-antidefect
is {\em not} screened in the IR through the static configuration.
The effective mobility from equation (\ref{EQN:MOBILITY}) is
$\eta \sim - \left[ \ln{\left| \xi dl/dt \right|} \right]^{-1}$
for a small defect pair, which agrees with  Ryskin and Kremenetsky
\cite{Ryskin91}. We see that for
non-zero velocities the mobility does not depend on the
system size, if the system is large enough.

We can estimate $\left< v^2 \right>_k$ for 2D XY systems with our results
in Table \ref{TAB:2DXYMICRO}. This will enable us to determine the UV
asymptotics of $T(k,t)$ directly through (\ref{EQN:TTLARGEK}), and with
that to determine the growth laws of 2D XY systems which scale
--- which we were unable to do in section \ref{sec:growth}.  We assume that
vortex/anti-vortex pairs with separation $l \ll L$ evolve
as described in Table  \ref{TAB:2DXYMICRO}, with an rms speed of annihilation
given by $v \sim dl/dt$. Then the number flux of defect pairs
is $j(l)=n(l)v(l)$.
If small features dominate then for $l \ll L$ we use $ j(l) \sim
j(\xi) \sim - \dot{L} L^{-(d+1)}$ in (\ref{EQN:VELSQUARE}), using $n=2$,
while if
small features do not dominate then the same approximation will give the
appropriate $\langle v^2 \rangle_k \sim \dot{L}^2$.
To evaluate the integral we take $v(l) \sim dl/dt$ for $l \ll L$
and use table \ref{TAB:2DXYMICRO}.  For 2D XY systems,
all vortex separations contribute to the energy dissipation since
$T({\bf k},l)$ has the same form for large and small $kl$.
However we require the quasi-adiabatic approximation to
apply for the smallest separation, so we need  $k > k_{cut}(l_{min})$.
This leads to $l_{\text{min}} \sim k^{-1}$.
We use $l_{\text{min}}$ in equation (\ref{EQN:VELSQUARE}) and
summarize our results in table \ref{TAB:SRVELSQUARE}. These results
can be used in equation (\ref{EQN:TTLARGEK})
to get the true $kL \gg 1$ limit of $T({\bf k},t)$.
We use these true asymptotics of $T({\bf k},t)$
in the energy-dissipation integral (\ref{EQN:ENDISS}). The energy-dissipation
integral is logarithmically divergent with
$\dot{\epsilon} \sim - \dot{L} L^{-3} \ln{(L/\xi)}$. Equating this
to the time derivative of $\epsilon$ from (\ref{EQN:E})
we find that, although scaling is
consistent, the growth law is not determined.
However, imposing the scaling form
(\ref{EQN:TWOTIMES}) on the asymptotics
(\ref{EQN:TTLARGEK}) of $T({\bf k},t)$, using table \ref{TAB:SRVELSQUARE},
determines the growth law, $L \sim (t/\ln{t})^{1/2}$, for non-conserved
systems and $L \sim t^{1/(2+\mu)}$ for conserved systems.
These growth laws differ by logarithmic factors in comparison to XY models
in $d>2$ (see Fig. \ref{FIG:SHORTRANGE}), so that the 2D XY model is a
special case of our both the treatment and the growth laws.

\subsection{Textures}
\label{sec:textures}

Systems with $n>d$ cannot have stable topological defects with singular
cores, but
systems with $n=d+1$ can support a non-singular topological
texture \cite{Toulouse76}. We construct an isolated texture of scale $X$ by
stereographically
projecting the field configuration from a $d$-sphere of radius $R= X/2$
surrounding an $n$-dimensional point defect in $n$-dimensional space
onto the $d$-dimensional space of the physical system.
For $d=1$ we rest a circle (1-sphere) on the 1D system. For $d=2$
we rest a sphere (2-sphere) on the planar system. We set $\vec{\phi}({\bf x})$
parallel to the radius vector of the sphere at the point of intersection
with the line joining ${\bf x}$ to the top of the sphere ($p_+$), as shown in
Fig. \ref{FIG:TEXTURE}. This constructs an $n$-component texture that winds
once over the $d$-dimensional system. The texture is topologically stable and
can only vanish when $X \rightarrow 0$. We obtain
\begin{equation}
\label{EQN:TEXTURE}
	\vec{\phi}({\bf x})= (p_{\bf x} \hat{x}, n_{\bf x}),
\end{equation}
where the $d$-sphere touches the system at $x=0$.
{}From Fig. \ref{FIG:TEXTURE} we have
\begin{eqnarray}
\label{EQN:DX}
p_{\bf x} &=& \sin{2 \alpha}, \nonumber \\
&=& 2 x X / (X^2+x^2),
\end{eqnarray}
and
\begin{eqnarray}
\label{EQN:NX}
n_{\bf x}&=& - \cos{2 \alpha}, \nonumber \\
 &=& (x^2-X^2)/(X^2+x^2).
\end{eqnarray}
The scale $X$ of the non-singular texture is determined by $n_{X}=0$.
We find
\begin{equation}
\label{EQN:TEXTCORR}
	\vec{\phi}({\bf x}) \cdot \vec{\phi}({\bf x }+{\bf r})
	= 1 - \frac{2 r^2 X^2 }{(X^2+x^2) \left[
		X^2 + ({\bf x}+ {\bf r})^2 \right] },
\end{equation}
and
\begin{eqnarray}
\label{EQN:TEXTSTRUCT}
	S({\bf k},X) &=&
	\rho_{\text{def}} \int d^d x \int d^d r e^{i {\bf k} \cdot {\bf r}}
	\vec{\phi}({\bf x}) \cdot \vec{\phi}({\bf x }+{\bf r}), \nonumber \\
	&=& (2 \pi)^d \delta ({\bf k})+2 \rho_{\text{def}} X^2
		{\bf \nabla}_{\bf k}^2
	\left[ \int d^d x \frac{e^{i {\bf k} \cdot {\bf x}}}{X^2+x^2}
\right]^2, \nonumber \\
	&\simeq&
2^{d+2} \pi^{d+1} \rho_{\text{def}}
		\frac{X^{d+1}}{k^{d-1}} e^{-2 k X}, \ \ \ \ \ \ \ \ k X \gg 1,
\end{eqnarray}
where $\rho_{\text{def}}$ is the number density of
defects (in the case of a single texture it is the inverse volume of the
system)
and we use the asymptotics of
\begin{equation}
	\int d^d x \frac{e^{i {\bf k}\cdot {\bf x}}}{X^2+x^2}
	= (2\pi)^{d/2} X^{d-2} (k X)^{1-d/2} K_{d/2-1}(k X),
\end{equation}
where $K_\nu(x)$ is a Bessel function of imaginary argument
 \cite{Gradshteyn65}.
For $d=2$ this construction generates
the minimal energy texture \cite{Belavin75} and has the same structure as
unpublished results of Bray and Puri \cite{Bray94b}.
In general dimension, this stereoscopic projection does not give the
minimal energy, and we expect that a different convex surface
(other than the $d$-sphere) is needed. However, we expect that the form of
the structure factor for $k X \gg 1$ will be unchanged and
that the scaling form for a {\em single} texture will be
\begin{equation}
\label{EQN:TEXTSCALE}
	S({\bf k},X) = X^{2d} g_T(kX),
\end{equation}
which follows directly from the second line of equation (\ref{EQN:TEXTSTRUCT}).
The time-derivative structure for $k X \gg 1$ is,
using Eqs. (\ref{EQN:TEXTURE}) to (\ref{EQN:NX}):
\begin{eqnarray}
\label{EQN:TTEXT}
	T({\bf k},X) &=& \rho_{\text{def}}\left. \dot{X}^2
			\frac{\partial^2}{\partial X_1 \partial X_2}
			\right|_{X_1=X_2=X}
\nonumber \\ &&
			\int d^d x \int d^d r
			\left[
		e^{i {\bf k} \cdot {\bf r}} \vec{\phi}({\bf x},X_1) \cdot
			\vec{\phi}({\bf x}+{\bf r},X_2) \right], \nonumber \\
&\simeq& 2^{d+2} \pi^{d+1} \frac{X^{d+1}}{ k^{d-3}} \dot{X}^2
	\rho_{\text{def}}	e^{-2 k X}, \ \ \ \ \ \ \ \ \ \ \ \ \ k X \gg 1.
\end{eqnarray}
This also leads to a scaling form for the time-derivative structure of
a {\em single} texture:
\begin{equation}
\label{EQN:TTTEXTSCALE}
 T({\bf k},X) = \dot{X}^2 X^{2d-2} h_T(kX).
\end{equation}

We use these scaling forms to calculate the energy of an isolated texture
of scale $X$. The energy will be, using
equations (\ref{EQN:EN}) and (\ref{EQN:TEXTSCALE}),
\begin{eqnarray}
\label{EQN:ETEXT}
	E(X) & = &
	V \int \frac{d^d k}{(2\pi)^d} k^2 S({\bf k},X),
	\nonumber \\
	& \sim & \int_0^\infty d^d k k^2  X^{2d} g_T(kX), \nonumber \\
	&\sim & X^{d-2},
\end{eqnarray}
where $V$ is the system volume, so $V \rho_{\text{def}} \sim 1$ for an isolated
texture. The rate of energy-dissipation is, using
equations (\ref{EQN:ENDISS}) and (\ref{EQN:TTTEXTSCALE}),
\begin{eqnarray}
\label{EQN:DISSTEXT}
	\dot{E}(X) &=& - V \int \frac{d^d k}{(2\pi)^d}
				T({\bf k},X), \nonumber \\
	&\sim & - \int_0^\infty d^d k \dot{X}^2 X^{2d-2} h_T(kX), \nonumber \\
	&\sim &   - \dot{X}^2 X^{d-2}.
\end{eqnarray}
Comparing these results, we see that
\begin{equation}
\label{EQN:DOTRTEXT}
	\dot{X} \sim - (d-2) X^{-1}.
\end{equation}
For $d>2$ an isolated texture will shrink
(in agreement with Derrick's theorem \cite{Derrick64}) and
vanish with $X(\tilde{t}) \sim \tilde{t}^{1/2}$ as a function of
time to annihilation, in $d=1$ isolated textures expand, while
2D textures are stable in this treatment.
Indeed, in 2D Belavin and Polyakov \cite{Belavin75}
present an exact solution for a static system
of textures, with $|\vec{\phi}|=1$ everywhere and
an energy independent of the scale
and position of the individual textures. However, systems with mixtures
of textures and anti-textures will be unstable \cite{Rutenberg94b}.

We can use these results to discuss the phase-ordering of systems with
many topological textures.  However we can go far with the
assumption that systems with textures do not have more singular correlations
than systems with point defects ($n=d$). With this,
we obtain from (\ref{EQN:POROD}) and (\ref{EQN:TTLARGEK}) that
$S({\bf k},t) \lesssim k^{-2d}$, and $T({\bf k},t)
\lesssim k^{2-2d}$ for $k L \gg 1$. We find that the integrals
(\ref{EQN:EN}) and (\ref{EQN:ENDISS}) converge for  systems with
textures if $d>2$. Hence for these textured systems $L \sim t^{1/(2+\mu)}$,
with or without conservation laws. This is in accord with the dimensional
argument of section \ref{sec:model}.

For non-conserved systems, we can use the results for isolated textures to
calculate the asymptotic $k L \gg 1$ structure factor due to a distribution
of textures at small scales. Very small textures, with $X \ll L$, will
collapse independently for $d>2$. The flux of annihilating textures
is $j(\xi) = - \dot{N}$, and the number density of textures scales
as an inverse volume $N \sim L^{-d}$. The flux of annihilating textures
at small scales, is $j(X)= n_t(X) \dot{X}$,
where $n_t(X)$ is the number density of textures at scale $X$.
Using $\dot{X} \sim X^{-1}$ from (\ref{EQN:DOTRTEXT}) and
$n_t(X) \sim \dot{N}/\dot{X}$ for $X \ll L$, we obtain
\begin{eqnarray}
\label{EQN:UVSTEXT}
	S({\bf k},t) &\simeq& \int_0^\infty
				dX \, n_t(X) S({\bf k},X), \nonumber \\
	     & \sim & 			\dot{N} k^{-(2d+2)},
		\ \ \ \ \ \ \ \ \ \ \ \ \ k L \gg 1, \ \ \ \ \ d>2,
\end{eqnarray}
and
\begin{eqnarray}
\label{EQN:UVTTTEXT}
	T({\bf k},t) &\simeq& \int_0^\infty dX \,
		n_t(X) T({\bf k},X), \nonumber \\
	     & \sim & \dot{N} k^{2-2d},
		\ \ \ \ \ \ \ \ \ \ \ \ \ \ \ k L \gg 1, \ \ \ \  d>2,
\end{eqnarray}
where we have used the scaling structure of an isolated texture
from equations (\ref{EQN:TEXTSCALE}) and (\ref{EQN:TTTEXTSCALE}).
These results satisfy the bounds we obtained from the assumption that
the structure will not be more singular than for systems with point
defects --- though the time-derivative correlations saturate the bounds.
The $k^{-8}$ tail predicted for the structure factor of three-dimensional
$n=4$ systems is in rough agreement with Toyoki's simulation results:
$S(k) \sim k^{-7.5 \pm 0.2}$ on the same system \cite{Toyoki94}.

For $d=1$ isolated textures will expand, because the energy of an
isolated texture of scale $X$ {\em decreases} with $X$ (\ref{EQN:DOTRTEXT}).
Hence we do not expect a singular annihilation process.
This will also be true in $d=1$ systems
with conservation laws. Rather, we expect a non-singular
combination of winding and anti-winding textures. The
minimal texture scale will increase with time, and the exponential factors
in (\ref{EQN:TEXTSTRUCT}) and (\ref{EQN:TTEXT}) will cause the
momentum integrals to converge in the UV. This
results in a {\em scaling} prediction of $L \sim t^{1/(2+\mu)}$.
In fact, it can be shown that both non-conserved and conserved 1D XY systems
do not scale \cite{Rutenberg94e}. The scaling violations are associated
with the fact that the correlation length $\xi_0$ for the initial conditions
enters in a nontrivial way. Extracting a length-scale from the
energy-density, $\epsilon \sim L^{-2}$, the non-conserved model
has $L \sim t^{1/4}$ \cite{Newman90a}, while the
conserved model has $L \sim t^{1/6}$ \cite{Mondello93,Rao94,Rutenberg94e}.
These growth laws are different than the scaling results, which is consistent
with the lack of scaling.

For $d=2$, textures will either slowly shrink due to an instability
caused by a soft-potential \cite{Rutenberg94b,Benson93}, or will unwind with
anti-textures in a non-singular annihilation process. In the former case,
which does not seem to be the dominant process \cite{Rutenberg94b}, a
similar approach to Eqs. (\ref{EQN:UVSTEXT}) and (\ref{EQN:UVTTTEXT}) finds
bounds which lead to convergent energy and energy-dissipation integrals
for any annihilation rate. In the latter case, with non-singular processes,
we also expect convergent integrals. In either case we expect
$L \sim t^{1/(2+\mu)}$ when scaling is obeyed --- however, different growth
laws and scaling violations are observed in simulations with non-conserved
dynamics \cite{Rutenberg94b}. Again, the scaling violations seem to be
associated with a nontrivial role of the length scale $\xi_0$ characterizing
the initial conditions \cite{Rutenberg94b}. Conserved 2D $n=3$ systems
have not yet been investigated.

The scaling assumption does not hold in the 1D and 2D systems with textures
that have been investigated. Does it hold for $d>2$?  The scaling
violations observed seem to be related to the weak interaction of the textures,
which is connected to the lack of long-range order above $T=0$ in equilibrium
correlations. Systems with textures in $d>2$ have low-temperature ordered
phases, and so are qualitatively different than 1D and 2D systems.
Indeed, we know of no scaling violations in systems with textures in $d>2$.

\section{Comparison with previous results}
\label{sec:review}

In this section we review the current understanding of growth laws,
and the evidence for our scaling assumptions, for various phase-ordering
systems. Other references can be found in reviews for scalar systems
\cite{Gunton83,Langer92,Furukawa85} and in the review of Bray \cite{Bray93e}.
The numerical simulations and experiments seem to be able to
check whether a particular growth law
(with or without logarithmic factor) is reasonable, or to
determine a growth-law exponent within about $10\%$ without
any theoretical input. The scaling of $T(k,t)$ is usually not considered.
Within these constraints, existing results are consistent with the
predictions of the Energy-Scaling approach, as we discuss in more detail
below.

\subsection{Non-conserved systems}
Scalar non-conserved systems, in agreement with our results, have
growth laws $L(t) \sim t^{1/2}$ as long as the growth
is driven by the curvature of domain walls, i.e.
for spatial dimension $d>1$ \cite{Allen79,Lifshitz62}.
This growth law has been confirmed
experimentally in two and three dimensions, see for example the work by Mason
{\em et al.} on
twisted nematic liquid crystals \cite{Mason93}, and the work by
Shannon {\em et al.} on $Cu_3Au$ \cite{Shannon92}.
In addition, simulations, such as Monte-Carlo (RG) studies \cite{Roland88},
also confirm this growth law. The scaling of the equal-time correlation
function (\ref{EQN:STRUCT}) also has extensive confirmation, for example
see the review by Furukawa \cite{Furukawa85}.
In one dimension, scalar systems exhibit growth driven by
exponentially suppressed interactions between domain walls. This case has been
treated approximately by Nagai and coworkers \cite{Nagai83} who found a
logarithmic growth law, and their solution and growth law can be shown
to be asymptotically correct \cite{Rutenberg94a}.

Vector non-conserved systems have not been as well understood theoretically.
We predict $L \sim t^{1/2}$ for all cases satisfying scaling, except
for the 2D XY model where we expect $L \sim (t/\ln{t})^{1/2}$.
This has been confirmed in the limit as
the number of components $n \rightarrow \infty$ \cite{Coniglio89,Pasquale83}.
The 1D XY model has two time-dependent length scales
\cite{Rutenberg94e}. A modified version of the Energy-Scaling treatment
may still be used, however, and the growth laws determined if an independent
relation between the length scales can be established by physical
arguments \cite{Rutenberg94e}. Experimentally, liquid-crystal XY-like
systems seem to exhibit $L \sim t^{1/2}$ in two \cite{Wong92} and
three \cite{Wong92,Chuang91} dimensions.

In 1D, simulations for systems without topological defects, with $n=3$, $4$,
and $5$, obtain a single length scale $L \sim t^{1/2}$, while
simulations for $n=2$ are consistent with the scaling violations mentioned
above \cite{Newman90a,Rutenberg94e}.

In 2D, careful simulations of XY models were consistent with
logarithmic factors, $L \sim (t/\ln t)^{1/2}$, in the growth law
\cite{Yurke93}.
These factors manifest themselves as a
systematic curvature in power-law fits of earlier numerical studies
\cite{Mondello90,Blundell92}. However,
tentative scaling violations in XY models have
been found by Blundell and Bray \cite{Blundell94}, when the length
scale was extracted from the defect density rather than directly
from the correlation function. Similar inconsistencies
are seen in simulations of uniaxial and biaxial
2D nematics ($n=2$) by Zapotocky {\em et al} \cite{Zapotocky94}.
No violations are seen when the correlations are collapsed with a length-scale
extracted from the correlation function itself (see e.g. \cite{Toyoki93}).
Measurement of $T({\bf k},t)$, the time-derivative correlations, may
help to resolve the scaling, or absence of scaling, of these systems.
For $n>3$ both hard spin ($|\vec{\phi}| \equiv 1$) \cite{Bray90} and
soft-spin \cite{Toyoki93} simulations seem to find $L \sim t^{1/2}$.
For $n=3$ the hard spin simulations find $L \sim t^{1/3}$ from the
energy density ($\epsilon \sim L^{-2}$), but no scaling \cite{Bray90}, which
is confirmed by more extensive simulations of both hard and soft-spin texture
systems \cite{Rutenberg94b}. [Recent work by Toyoki \cite{Toyoki93}, which
claimed scaling and $L \sim t^{1/2}$ in this system, is hampered by early-time
transients.]

In three dimensions, a growth law
$L \sim t^{0.45 \pm 0.01}$ has been observed for the XY model
\cite{Blundell92,Blundell94,Toyoki91a,Mondello92} ---
less than the $t^{1/2}$ prediction.  However the system
seems to scale \cite{Blundell94}.  The suppressed growth law
may be related to pinning effects due to the discrete dynamics
--- a possible explanation \cite{Blundell94,Toyoki93}
for anomalously small growth exponents
seen in simulations done at $T=0$.  Alternatively, corrections to
scaling may be responsible.  Simulations by Toyoki
found $L \sim t^{1/2}$  for 3D Heisenberg ($n=3$)
systems \cite{Toyoki91b}. Scaling is confirmed by Blundell and
Bray \cite{Blundell94}, though again
they obtained growth exponents slightly less than $1/2$.

Our results derived from the scaling assumption are consistent with the bulk
of the evidence. Further work needs to be done to
measure $T({\bf k},t)$ in 2D XY
systems to resolve the issue of scaling, and to check the
deviations from $t^{1/2}$ growth seen in $d>2$ vector systems.

\subsection{Conserved systems}

We confirm that scalar conserved systems with $\mu=2$ have the standard growth
law of $L \sim t^{1/3}$ \cite{Lifshitz61,Huse86},
provided the spatial dimension $d >1$ so that the
evolution of domain walls is curvature driven.  This result is also
obtained by RG arguments \cite{Bray89} and is seen experimentally,
for instance in the binary alloy $Mn_{0.67}Cu_{0.33}$
\cite{Gaulin87}, as well as in simulations
\cite{Roland88,Huse86}. As in the non-conserved case, scaling of
$S({\bf k},t)$ has been confirmed
\cite{Furukawa85} but the scaling of $T({\bf k},t)$
has not been considered.
Kawakatsu and Munakata \cite{Kawakatsu85}
have considered the 1D case, in which there are no curvature
effects, and found logarithmic growth in agreement with our results
in section \ref{sec:ediss}.
Scalar models with generalized conservation laws ($\mu>0$) have
also been considered, the so called ``noninteger derivative'' or
``long-range exchange'' models \cite{Onuki85a}.
For scalar systems, growth laws in agreement with our results
are found, both through computer simulations and by considering
the evolution of small defect features.

For vector conserved systems, with $\mu=2$, we predict growth
laws of $L \sim t^{1/4}$ --- except for XY systems with $d>2$ where we
expect $L \sim (t \ln{t})^{1/4}$.
This agrees with RG arguments, which, assuming standard scaling,
found growth exponents of $1/4$ for all $n \geq 2$ \cite{Bray89} (we note
again that the RG treatment does not determine logarithmic factors).
Coniglio and Zannetti found scaling violations for $n=\infty$
\cite{Coniglio89}, so that our treatment does not apply to that case.
However, Bray and Humayun \cite{Bray92a} show
how standard scaling could be recovered for any finite $n$.

Simulations of the conserved 1D XY model find a scaling violation
that is consistent with the discussion
at the end of section \ref{sec:textures} \cite{Rutenberg94e}.
Simulations of the 2D XY model \cite{Mondello93} obtain $L \sim t^{1/4}$,
though small but systematic scaling violations seem to be indicated by
the data. These ``violations'' could just be strong corrections to scaling,
related to the corrections of order $O(1/\ln{l})$ expected in the growth-law
(from the cancelled logarithm from the energy-density in (\ref{EQN:E})).
Recent simulations by Puri et al.\ \cite{Puri94} found good scaling for
both $d=2$ and $d=3$, with growth laws consistent with the forms
$L \sim t^{1/4}$ ($d=2$) and $L \sim (t \ln t)^{1/4}$ ($d=3$) predicted here.
For the three-dimensional XY model, earlier work by Siegert and Rao
\cite{Siegert93} also found good evidence for standard scaling. Although
the growth law was originally interpreted as $L\sim t^{0.28}$
\cite{Siegert93}, the data are much better fitted by the Energy-Scaling
prediction $L \sim (t \ln t)^{1/4}$ \cite{Siegert94}.

Again, our results are consistent with previous work,
particularly the RG results for the power-law factors in the growth
laws. Conserved 2D systems with textures ($n=3$) should be investigated.
In addition, the possible scaling violations in the conserved
2D XY system need to be resolved,  perhaps by measuring $T({\bf k},t)$.

\subsection{Long-range systems}

For systems with attractive
long-range interactions (\ref{EQN:LONGRANGE}),
Bray \cite{Bray93b} treated the conserved case with an RG
approach, and the non-conserved scalar case with physical arguments.
Our results are in agreement, with additional logarithmic factors in
marginal cases,
for scalar and vector systems for $0<\sigma<2$. Numerical work by
Hayakawa {\em et al.} \cite{Hayakawa93a} for non-conserved 2D scalar systems
roughly agree with our results for $\sigma=0.5$, $1.0$, and $1.5$, but
more sensitivity is needed to test the logarithmic factor for $\sigma= 1$.
Analytical work by Hayakawa {\em et al.} \cite{Hayakawa93b}
found a breaking of scaling for conserved dynamics in the spherical limit,
similar to that seen in the short-range case \cite{Coniglio89}.
Scaling violations in {\em non-conserved}
1D scalar systems for $\sigma<1$ have been proposed by Lee and Cardy
\cite{Lee93}. However, no scaling violations in either $S({\bf k},t)$ or
$T({\bf k},t)$ are seen in more
extensive simulations of 1D scalar systems \cite{Rutenberg94a}, which
find growth laws in agreement with our predictions for $\sigma=0.5$,
$1.0$, and $1.5$.

\section{Discussion}
\label{sec:discussion}

Our approach does not depend on the initial conditions of the system
\cite{nodefect}.
If scaling is obeyed, critical ($ \left< \vec{\phi} \right> = 0$)
and off-critical ($ \left< \vec{\phi} \right> \neq 0$)
quenches will have identical growth laws. For non-conserved
systems, off-critical quenches are found to break scaling
\cite{Toyoki87,Bray92c}, since the order-parameter saturates,
and so our growth laws do not apply. Indeed,
exponentially growing length-scales are seen \cite{Mondello92}.
Conserved systems, on the other hand, have the same growth law for
critical and off critical quenches, in agreement with earlier predictions
\cite{Huse86,Bray89}. This has been numerically confirmed
in the scalar case \cite{Chakrabarti93}.
Initial conditions with long-range correlations (corresponding to, e.g.,
quenches from a critical point) will, similarly,
have the same growth laws if scaling holds \cite{Bray91b}.
In general, various classes of initial conditions may affect the
existence of scaling and the
form of the scaling functions (\ref{EQN:STRUCT}) and (\ref{EQN:TWOTIMES}),
but will not change the growth laws if scaling holds.
Of course, the appropriate defect structure must be applied. For example,
if the 2D XY model is quenched {\em from} below $T_{KT}$
then the bound vortex pairs quickly annihilate and the scaling configurations
have no topological defects (for a quench to $T=0$). In this case scaling
is obeyed and the growth law is $L \sim t^{1/2}$ \cite{Rutenberg94d},
as expected for a defect-free quench, without the logarithmic factor of
2D XY systems quenched from above $T_{KT}$.

We derive our results at zero-temperature but we expect our growth laws
to describe quenches to all temperatures below $T_c$. We do not
address quenches in which thermal noise
is essential, such as systems with static disorder \cite{Huse85},
or quenches to a $T>0$ critical point \cite{Janssen89}.
For systems with long-range order only at $T=0$, such as
XY and scalar systems in one-dimension,
our approach applies only at $T=0$. For small $T$, such that the equilibrium
correlation length $\xi(T)$ of the disordered phase is large, we expect
our growth laws to apply as long as $L(t) \ll \xi(T)$.
The 2D XY model quenched from high-temperature to $0 < T < T_{KT}$
develops power-law order and invites further study because the
low-temperature behavior is not described by a single $T=0$ fixed point
but by a line of fixed points.
Numerical work for this case by Yurke {\em et al.}
\cite{Yurke93}, at $0<T<T_{KT}$, found a growth law in agreement with our
$T=0$ result, suggesting a similar growth-law for all temperatures below
$T_{KT}$.

Our treatment does not depend on the details of the potential $V(\vec{\phi})$
in the energy-functional.  It is only the symmetry properties
of the potential's ground-state manifold that determine the
growth law, since the ground-state manifold determines the defect
structure, which in turn determines the asymptotic structure at $kL \gg 1$
through simple scaling arguments. The Energy-Scaling
approach can be applied to systems with more complicated order
parameters than $n$-component vectors. All we need is the existence of some
short or long-ranged ``elastic'' energy ($\sigma$), a conservation
law ($\mu$), and the defect structure, if any.
The effective $n$ corresponding to a defect type is the one that determines
the scaling of its core volume density $\rho_{\text{core}} \sim L^{-n}$.
The defect type with the smallest $n$, and hence greatest core volume,
will determine the growth law, since
it will provide the dominant contribution
to any UV divergences in the energy density or dissipation integrals.
Systems with non-abelian symmetries, such as cholesteric liquid crystals,
 are treated in the same way, since
our approach is predicated on the energetics of the system rather than
on the detailed nature of the dynamics.
This assumes that scaling holds, so that {\em all}
defect types scale with the same growth law.

For example, in bulk uniaxial or biaxial
nematic liquid crystals the existence of string
defects, with $\rho_{\text{string}} \sim L^{-2}$, leads to $n=2$.
Using this in equations (\ref{EQN:POROD}) and (\ref{EQN:TTLARGEK})
with no conservation law implies a $L \sim t^{1/2}$ growth law. This is
consistent with recent
experiments \cite{Wong92,Chuang91} and simulations \cite{Blundell92}.
We neglect the point defects, since they do not dominate
the asymptotics or the energetics.

Similarly, in Potts models the existence of domain walls leads to
the same growth laws as for scalar ($n=1$) systems, consistent with
the $L \sim t^{1/2}$ growth seen in non-conserved systems \cite{Roland90},
and the $L \sim t^{1/3}$ growth suggested by numerical studies of
conserved systems \cite{Jeppesen93}.
The existence of vertices in Potts models,
or e.g. in clock models \cite{Kawasaki85}, will not change the form of the
asymptotic growth law. At late times the energy density
and dissipation will be dominated by the change of domain wall volume,
with $\rho_{\text{wall}} \sim L^{-1}$,  rather than by the
vertices, with $\rho_{\text{vertices}} \sim L^{-2}$. In general,
the prefactor or timescale
of the growth law will depend on the details of the model, but the
exponent and any logarithmic factors of the growth law will be universal.

Our approach is restricted to systems which
are governed by dissipative dynamics and a simple energy-functional
of the form (\ref{EQN:HAMILTONIAN}) or (\ref{EQN:LONGRANGE}).
It would be interesting to speculate on the growth laws in the late stages
of defect elimination in patterned systems, with {\em competing} short and
long-range interactions. However, while the defect types of the patterned
structures (disclinations etc...)
can be clearly identified, it is not at all clear what conservation
laws or long-range forces apply to the effective order parameter of
the patterned structure, or even
whether temperature is relevant to phase-ordering \cite{Bahiana90}.
This is a promising direction for further research.

For the systems where we know the ``collapse'' laws $l(\tilde{t})$, of the
size of an isolated defect for a given time to collapse, they are the same
form as the scaling growth law $L(t)$, of the
phase-ordering length-scale for a given time after the quench.
This agrees with our naive expectation for scaling systems that the form of
the collapse law will be the same as the growth law --- the intuitive picture
is of collapsing defect features leaving ``voids'' which set the growing
length-scale, $L(t)$.

We can say little about the growth laws in systems that break our scaling
assumptions, apart from our discussion about the $\xi_0$ dependence at the
end of section \ref{sec:model}.
Some systems can be explicitly shown to break scaling,
such as 1D XY models \cite{Rutenberg94e},
or conserved spherical systems \cite{Coniglio89,Hayakawa93b}.
Are there other systems that break scaling?
We cannot answer this by examining only the equal-time
correlation function, since the scaling of the latter does {\em not} imply
the scaling of $T({\bf k},t)$. There are two paths to take. The first is to
explicitly find scaling violations in a system.  A possible example is given
by the equal-time correlations in the 2D XY model,
both for the non-conserved \cite{Blundell94,Zapotocky94} and for the conserved
(apparent in figure 3 of \cite{Mondello93}) cases.  The second method
is to find a growth law that is in striking disagreement with the
our predictions, such as in the 2D $n=3$ system \cite{Rutenberg94b}.
Since our approach is based on the scaling assumption,
any disagreement implies a scaling violation.
Conversely, to demonstrate scaling the correlations must be measured directly
since agreement between the observed and predicted growth laws
is necessary but not sufficient to demonstrate scaling.

\section{Summary}
\label{sec:summary}

In summary, by focusing on the energetics, rather than
the detailed dynamics of the system, we obtain growth laws for phase ordering
( summarized in figures \ref{FIG:SHORTRANGE} and \ref{FIG:LONGRANGE},
as well as table \ref{TAB:2DXYGROWTH}). This leads to a powerful,
unified, and physical approach to determining growth laws that
rests solely on the existence of scaling.  We call this the
``Energy-Scaling'' approach. It can be used for any system with purely
dissipative dynamics. Any disagreement of a growth law
from our predictions indicates the breaking of scaling
of at least one of the two-point correlation functions.
In particular, we stress the importance of the time-derivative
correlation function.

In addition to the growth laws, we determine the
collapse law for isolated non-conserved textures ($n=d+1$)
in more than two-dimensions, and use that
to determine the generalized Porod's law (\ref{EQN:UVSTEXT}) and asymptotics
for $T({\bf k},t)$ (\ref{EQN:UVTTTEXT})
--- these follow novel power-laws due to annihilating textures.
We also treat
the dynamics of isolated defects in scalar systems and 2D XY systems.
When scaling holds, we find that the growth law matches the form set by
collapsing isolated defect features, of the characteristic scale $L$, within
a larger phase-ordering system.

\acknowledgments

We thank T. Blum, R. E. Blundell, J. Cardy, D. A. Huse, B. P. Lee,
M. A. Moore, S. Puri, M. Siegert and M. Zapotocky for discussions.
We also thank J. Cardy, H. Hayakawa, M. Mondello, and H. Toyoki for
their correspondence, and the Isaac Newton Institute, Cambridge, where
this work was finished, for its hospitality. This work was supported
by EPSRC grant GR J82041.

\begin{figure}
\caption{The growth law $L(t)$ of the length scale for phase-ordering
systems which scale.
The growth laws are shown for various number of components $n$, where
$n \leq d$, and conservation laws, $\mu$.
The black circles correspond to possible physical systems.
For $n=d=1$ and $n=d=2$ the growth-laws in
table \protect\ref{TAB:2DXYGROWTH} apply.
Scaling systems with no topological defects ($n>d+1$), or
with textures ($n=d+1$), will have $L \sim t^{1/(2+\mu)}$.
\label{FIG:SHORTRANGE}}
\end{figure}

\begin{figure}
\caption{The growth law of the
length scale $L(t)$ for attractive long-range interactions with a fixed sigma,
$0<\sigma \leq 2$ for systems
which satisfy our scaling assumptions.
The growth laws are shown for
various number of components $n$
and  conservation laws $\mu$.
For $n=d=1 \leq \sigma <2$, $L(t) \sim t^{1/(1+\sigma)}$
\protect\cite{Rutenberg94a}.  For $\sigma=2$ and $n=d \leq 2$,
see table \protect\ref{TAB:2DXYGROWTH}.
Scaling systems with no topological defects ($n>d+1$), or
with textures ($n=d+1$), will have $L \sim t^{1/(\sigma+\mu)}$.
\label{FIG:LONGRANGE}}
\end{figure}

\begin{figure}
\caption{The construction of a $d+1$ component texture in a
$d$-dimensional system by stereographically projecting a $d$-sphere onto the
system (shown is a plane containing ${\bf x}$ and the
two-poles $p_{\pm}$ of the $d$-sphere).  The order-parameter
$\vec{\phi}({\bf x})=\hat{R}$, where the radial vector ${\bf R}$
meets the $d$-sphere at the intersection with a line joining ${\bf x}$ with
the pole $p_{+}$ of the sphere.
The scale of the constructed texture, $X$, is twice the
radius of the $d$-sphere.}
\label{FIG:TEXTURE}
\end{figure}

\begin{table}
\caption{Results of the energy dissipation and the domain size dynamics
of an isolated scalar domain of scale
$l \ll L $ with $d>1$. The time
left before domain annihilation is $\tilde{t}$.
\label{TAB:SCALAR}}
\end{table}

\begin{table}
\caption{Evolution of a pair of defects
of separation $l \ll L$ for $n=d=2$.
\label{TAB:2DXYMICRO}}
\end{table}

\begin{table}
\caption{The average square-velocity, $\left< v^2 \right>_k$, of
defect cores which contribute to $T({\bf k},t)$ at $kL \gg 1$ for scalar
and 2D XY systems. The dots indicate time derivatives.
\label{TAB:SRVELSQUARE}}
\end{table}

\begin{table}
\caption{Growth laws, $L(t)$,
for short-range systems with $n=d \leq 2$.
\label{TAB:2DXYGROWTH}}
\end{table}

\onecolumn

\begin{figure}
\begin{picture}(300,300)
\thicklines
\put(20,20){\vector(1,0){250}}
\put(20,20){\vector(0,1){250}}
\thinlines
\put(20,170){\line(1,-1){150}}
\put(20,170){\vector(1,0){250}}
\put(5,5){0}
\put(5,95){1}
\put(5,170){2}
\put(5,245){3}
\put(170,5){2}
\put(5,270){$n$}
\put(280,5){$\mu$}
\put(20,95){\circle*{5}}
\put(20,170){\circle*{5}}
\put(20,245){\circle*{5}}
\put(170,95){\circle*{5}}
\put(170,170){\circle*{5}}
\put(170,245){\circle*{5}}
\put(50,75){$t^{1/2}$}
\put(190,75){$t^{1/{(n+\mu)}}$}
\put(190,200){$t^{1/(2+\mu)}$}
\put(51,187){\vector(-2,-1){25}}
\put(52,188){$t^{1/2}$}
\put(77,154){\vector(-2,-1){25}}
\put(78,155){$(t/\ln{t})^{1/2}$}
\put(236,155){\vector(-2,1){25}}
\put(237,153){$(t \ln{t})^{1/(2+\mu)}$}
\end{picture}
\end{figure}

\begin{figure}
\begin{picture}(300,300)
\thicklines
\put(20,20){\vector(1,0){250}}
\put(20,20){\vector(0,1){250}}
\thinlines
\put(20,170){\line(1,-1){150}}
\put(20,50){\vector(1,0){250}}
\put(5,5){0}
\put(5,50){$\sigma$}
\put(5,95){1}
\put(5,170){2}
\put(5,245){3}
\put(170,5){2}
\put(5,270){$n$}
\put(280,5){$\mu$}
\put(20,95){\circle*{5}}
\put(20,170){\circle*{5}}
\put(20,245){\circle*{5}}
\put(170,95){\circle*{5}}
\put(170,170){\circle*{5}}
\put(170,245){\circle*{5}}
\put(35,100){$t^{1/(2+\sigma-n)}$}
\put(220,35){$t^{1/{(n+\mu)}}$}
\put(220,100){$t^{1/(\sigma+\mu)}$}
\put(35,35){$t^{1/2}$}
\put(75,64){\vector(2,-1){25}}
\put(48,68){$(t \ln{t})^{1/2}$}
\put(129,34){\vector(1,0){25}}
\put(82,34){$(t/\ln{t})^{1/2}$}
\put(166,64){\vector(-2,-1){25}}
\put(168,65){$t^{1/2}$}
\put(220,64){\vector(-2,-1){25}}
\put(221,65){$(t \ln{t})^{1/(n+\mu)}$}
\put(86,144){\vector(-2,-1){25}}
\put(88,145){$(t/\ln{t})^{1/(\sigma+\mu)}$}
\end{picture}
\end{figure}

\begin{figure}[h]
\begin{center}
\mbox{
\epsfxsize=5.0in
\epsfbox{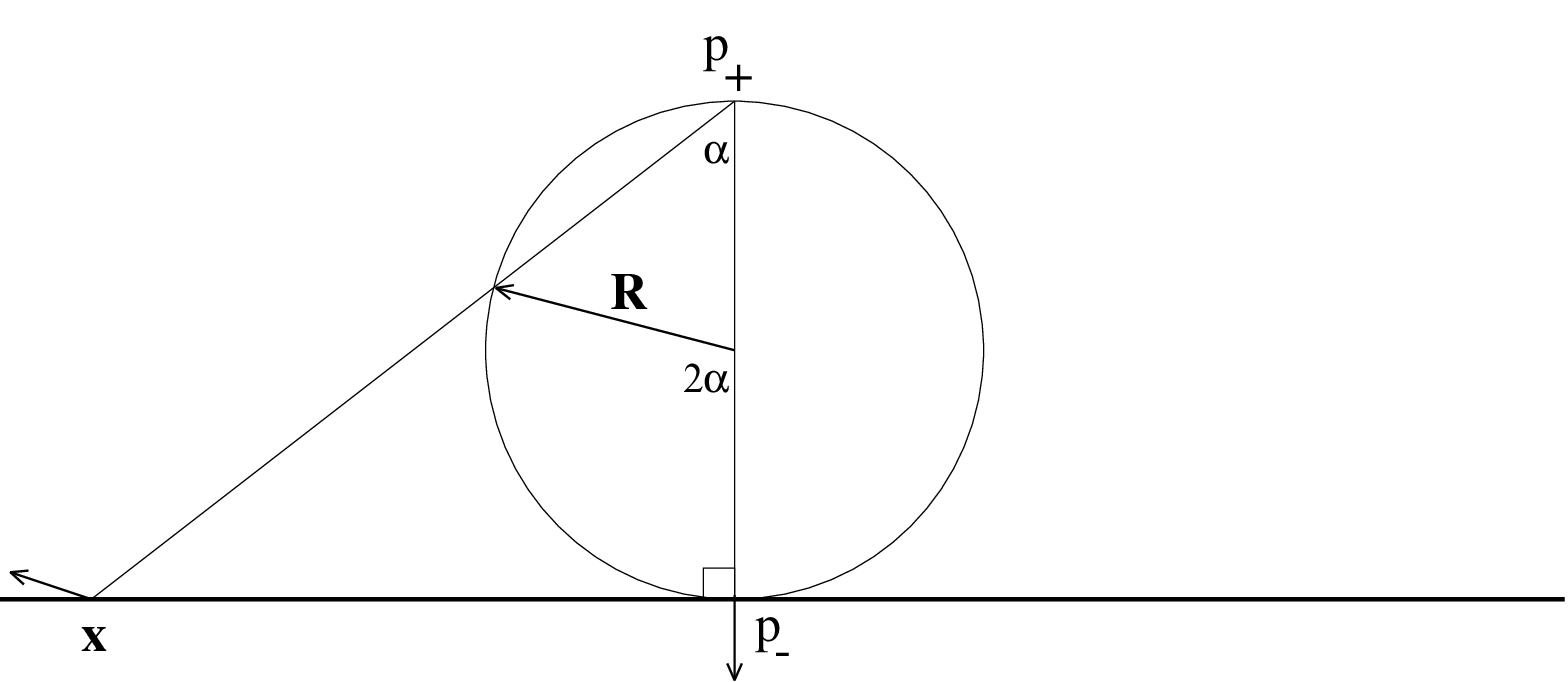}
}
\end{center}
\end{figure}

\begin{table}
\begin{tabular}{|l|ccc|} \hline
 	& $\partial_t{E}(l)$ & $dl/dt$ & $l(\tilde{t})$ \\ \hline
 $\mu < 1 $ 	& $- (dl/dt)^2 l^{d-1} \xi^{\mu-1}$
		& $ -l^{-1}$
		& $\tilde{t}^{1/2}$ \\
 $\mu =1 $ 	& $- (dl/dt)^2 l^{d-1} \ln{(l/\xi)}$
		& $ -\left[ l \ln{(l/\xi)} \right]^{-1}$
		& $(\tilde{t}/\ln{\tilde{t}})^{1/2}$ \\
 $1 < \mu <d $ 	& $-(dl/dt)^2 l^{d+\mu-2}$
		& $- l^{-\mu}$
		& $\tilde{t}^{1/(1+\mu)}$ \\
 $ 1 < \mu =d$	& $-(dl/dt)^2 l^{2d-2} \ln{(L/l)}$
		& $ - l^{-d}/\ln{(L/l)}$
		& $\left[\tilde{t}/\ln(L^{1+d}/\tilde{t})
			\right]^{1/(d+1)}$ \\
 $ \mu > d > 1$ & $-(dl/dt)^2 l^{2d-2} L^{\mu-d}$
		& $ - l^{-d} L^{d-\mu}$
		& $\tilde{t}^{1/(d+1)} L^{(d-\mu)/(d+1)}$ \\
\hline
\end{tabular}
\end{table}

\begin{table}
\begin{tabular}{|l|ccc|} \hline
 	& $\partial_t{E}(l)$ & $dl/dt$	& $l(\tilde{t})$ \\
\hline
 $\mu =0 $ 	& $(dl/dt)^2 \ln{\left| \xi dl/dt \right|} $
		& $\left[ l \ln{\left| \xi dl/dt \right|} \right]^{-1}$
		& $(\tilde{t}/\ln{\tilde{t}})^{1/2}$ \\
 $\mu >0 $ 	&$-(dl/dt)^{(2+\mu)/(1+\mu)}$
		& $-l^{-(1+\mu)}$
		& $\tilde{t}^{1/(2+\mu)}$ \\
\hline
\end{tabular}
\end{table}

\begin{table}
\begin{tabular}{|l|cc|} \hline
 	& $n=1<d$ & $n=d=2$ 	\\ \hline
$\mu=0$  	& $ \dot{L}^2$
		& $\dot{L} L^{-1} \ln(kL) / \ln(L/\xi) $ \\
$0< \mu < d$ 	& $ \dot{L}^2$
		& $\dot{L} L^{-1} k^\mu$   \\
$\mu = d$	& $\dot{L} L^{-\mu} \ln \ln{(kL)}$
		& $\dot{L} L^{-1} k^\mu$ \\
$\mu > d$	& $\dot{L} L^{-\mu} \ln(kL)$
		& $\dot{L} L^{-1} k^\mu$ \\
\hline
\end{tabular}
\end{table}

\begin{table}
\begin{tabular}{|l|cc|} \hline
 	& $n=d=1$ & $n=d=2$ 	\\ \hline
$\mu=0$  	& $ \ln{t}$
		& $(t/\ln{t})^{1/2}$ \\
$\mu >0$ 	& $ \ln{t}$
		& $t^{1/(2+\mu)}$   \\
\hline
\end{tabular}
\end{table}


\begin{references}
\bibitem{magnetic} We use magnetic language throughout, but our
results are not limited to magnetic systems.
\bibitem{MC} Monte Carlo RG approaches \protect\cite{Roland88} are
not restricted to conserved systems.
However, rather than predicting a particular growth law, they
select the trial growth law that best reduces the data within an
assumed fixed point structure.
\bibitem{ignore} We have assumed power-law growth (with or without logarithmic
factors) to obtain the time-derivative scaling form from (\ref{EQN:TTSTRUCT}).
\bibitem{feature} The number density $n(l)$ is defined by the total core
volume density with a local length-scale $l$ divided by the expected core
volume at that scale, $l^{d-n}$. For point defects ($n=d$), $n(l)$ is
the coarse-grained defect density distribution.
\bibitem{mobility}
For conserved scalar systems, a local ``effective mobility'' makes no sense,
since the local movement of a domain wall changes the average order
parameter.
\bibitem{metastable} Equation (\ref{EQN:2DXYFIELD}) is an analytic
solution to $\nabla^2 \theta =0$ and hence represents a metastable
configuration.  The finite core-size $\xi$, due to a finite $V_0$ or a
discrete lattice, destabilizes the configuration and leads to evolution.
The finite rate of evolution then causes (\ref{EQN:2DXYFIELD}) to break
down at large distances from the vortices.
\bibitem{nodefect} If there are no topological defects seeded by the initial
conditions, then we expect $L(t) \sim t^{1/(\sigma+\mu)}$
for scaling systems.
\end{references}
\end{document}